\documentclass[structabstract]{aa}  
\usepackage{graphicx}
\usepackage{txfonts}
\begin{document}
  \title{Effects of quark matter nucleation \\ 
on the evolution of proto--neutron stars}
%%   \subtitle{ }

   \author{Ignazio Bombaci 
          \inst{1, 2}\fnmsep\thanks{corresponding author. email: bombaci@df.unipi.it}
          \and
          Domenico Logoteta\inst{3}
          \and  
          Constan\c{c}a Provid\^encia\inst{3}
          \and
          Isaac Vida\~na\inst{3}  
          }

   \institute{Dipartimento di Fisica ``E. Fermi'', Universit\`a di Pisa, 
                  Largo B. Pontecorvo, 3 I-56127 Pisa, Italy\\
%%%%% \email{bombaci@df.unipi.it}
    \and 
                 INFN, Sezione di Pisa, Largo B. Pontecorvo, 3 I-56127 Pisa, Italy\\            
    \and
                 Centro de F\'{\i}sica Computacional, Department of Physics, University of Coimbra, 
                 3004-516  Coimbra, Portugal \\
             }

   \date{Received 18 September 2010;  accepted 03 December 2010}

% \abstract{}{}{}{}{} 
% 5 {} token are mandatory
 \abstract 
{A phase of strong interacting matter with deconfined quarks is expected in the core of a massive 
neutron star. If this deconfinement phase transition is of the first order, as suggested  by many models  
inspired by quantum chromodynamics, then it will be triggered by the  nucleation of a critical size drop 
of the (stable) quark  phase in the metastable hadronic phase.  
Within these circumstances it has been shown that cold ($T=0$) pure hadronic compact stars above 
a threshold value of their gravitational mass (central pressure) are metastable with respect to  the ``decay'' (conversion) to  quark stars ({\it i.e.}, compact stars made at least in part of quark matter). 
This stellar conversion process liberates a huge amount of energy (a few $10^{53}$ erg),  and it could 
be the energy source of some of the long gamma ray bursts. }
{The main goal of the present work is to establish whether a newborn hadronic star (proto-hadronic star) 
could survive the early stages of its evolution without ``decaying" to a quark star.   
To this aim, we study the nucleation process of quark matter in hot ($T \neq 0$) 
$\beta$-stable hadronic matter, with and without trapped neutrinos, using a finite temperature 
equation of state (EOS) for hadronic and quark matter. }
{The finite--temperature EOS for the hadronic and for the quark phases were calculated using 
the nonlinear Walecka model and the MIT bag model, respectively.  
The quantum nucleation rate was calculated making use of the Lifshitz \& Kagan nucleation theory. 
The thermal nucleation rate was calculated using the Langer nucleation theory.}
{We calculate and compare the nucleation rate and the nucleation time due to thermal  and quantum nucleation mechanisms. We compute the crossover temperature above which thermal nucleation 
dominates the finite temperature quantum nucleation mechanism. 
We next discuss the consequences of quark matter nucleation for the physics and the evolution 
of proto-neutron stars. 
We introduce the new concept of limiting conversion temperature  and 
critical mass  $M_{cr}$ for proto-hadronic stars, and we show that  proto-hadronic stars with a 
mass $M < M_{cr}$ could survive the early stages of their evolution without decaying to a quark star.  
We extend the concept of maximum mass of a ``neutron star'' with respect to the classical one 
introduced by Oppenheimer \& Volkoff to account for the existence of two 
distinct families of compact stars (hadronic stars and quark stars) as predicted by the 
present scenario.}
   {}
%---------------------------------------------------
\authorrunning{Bombaci et al.}
\titlerunning{Effects of quark matter nucleation on the evolution of proto--neutron stars}

  \maketitle
%
%________________________________________________________________
\section{Introduction}

According to quantum chromodynamics (QCD) a phase transition from hadronic matter to a 
deconfined quark phase should  occur at a density of a few times nuclear matter saturation 
density $\rho_0 \sim 2.8 \times 10^{14}$~g/cm$^{3}$.   
Neutron star structure calculations, based on a wide variety of modern equations of state (EOS) 
of hadronic matter (Lattimer \& Prakash \cite{LP01}; Bombaci \cite{bomb07}),  predict a 
maximum stellar central density (the one for the maximum  mass star configuration) in the 
range of  4 -- 8 times $\rho_0$.  Consequently, the core of neutron stars is one of the best candidates in the universe where a phase of strong interacting matter with deconfined 
{\it up}, {\it down } and {\it strange} quarks could be  found.    

In the region of high density (high baryon chemical potential) and low temperature (which is the 
one relevant for neutron star physics) many QCD-inspired models suggest the deconfinement transition 
to be a first-order phase transition (Hsu \& Schwetz \cite{hs98}, Fodor \& Katz \cite{fk04}). 
As  is well known, first-order phase transitions are triggered by the  nucleation of a  
critical size drop of the new (stable) phase in a metastable mother phase. 

In recent years there has been a growing interest in studying the nucleation 
process of quark matter in the core of massive neutron stars.  
The nucleation rate of quark matter in cold ($T = 0$) $\beta$-stable nuclear  
matter has been calculated by Iida \& Sato (\cite{iida97}) wthin the Lifshitz--Kagan quantum  nucleation theory (Lifshitz \& Kagan \cite{lk72}).  The same authors (Iida \& Sato \cite{iida98}) 
have later on studied quark matter nucleation in $\beta$-stable  hyperonic matter within a relativistic extension of the Lifshitz--Kagan theory.     
Various significant astrophysical implications of quark matter nucleation in neutron stars 
have been explored in a series of subsequent papers 
(Berezhiani et al. \cite{be02}; \cite{be03}; Bombaci et al. \cite{bo04}; 
Drago et al. \cite{drago04}; Lugones \& Bombaci \cite{lug05}; Bombaci et al. \cite{blv07}; 
Bombaci et al. \cite{bppv08}; Drago et al. \cite{dps-b08}; Bambi \& Drago \cite{bambi08}).  
In particular, in these studies it has been shown that,  above a threshold value of the central pressure,  
a pure hadronic compact star (HS) is metastable to the decay (conversion) 
to a quark star (QS), {\it i.e.} to a hybrid neutron star or to a strange star 
(Bodmer \cite{bod71}; Witten \cite{witt84}), depending on the details of the EOS for 
quark matter used to model the phase transition.  
This stellar conversion process releases (Bombaci \& Datta \cite{grb}) an extraordinarily large 
amount of energy (a few $10^{53}$ erg) and it could be the energy source of some of the 
long gamma ray bursts (GRBs).       

The research reported in these papers has focused on the quark deconfinement 
phase transition in cold  and neutrino-free neutron stars. In this case the formation 
of the first drop of QM could take place solely via a quantum nucleation process. 

A neutron star at birth (proto-neutron star) is very hot (T =  10 -- 30 MeV) with neutrinos being 
still trapped in the stellar interior (Burrow \& Lattimer \cite{BurLat86}; Prakash et al. \cite{prak97}; Pons et al. \cite{pons99}).   
Subsequent neutrino diffusion causes deleptonization and heats the stellar matter to an approximately uniform entropy per baryon $\tilde {S}$ =1 -- 2 (in units of the Boltzmann's constant $k_B$).  
Depending on the stellar composition, during this stage neutrino escape can lead the more 
``massive'' stellar configurations to the formation of a black hole (Bombaci, \cite{bomb96}; 
Prakash et al. \cite{prak97}).  However, if the mass of the star is sufficiently low, 
the star will remain stable  and  will cool to temperatures well below 1~MeV  within 
a cooling time $t_{cool} \sim$~a~few~$10^2$~s, as the neutrinos continue to carry energy away 
from the stellar material (Burrow \& Lattimer \cite{BurLat86}; Prakash et al. \cite{prak97};
Pons et al. \cite{pons99}).    
Thus in a proto-neutron star, the quark deconfinement phase transition will likely be triggered 
by a  thermal nucleation process. In fact, for sufficiently high temperatures, thermal nucleation 
is a much more efficient process than the quantum nucleation mechanism.  

Some of the earlier studies of quark matter nucleation 
(see {\it e.g.}, Horvath et al. \cite{ho92}; Horvath \cite{ho94}; Olesen \& Madsen  
\cite{ol94}; Heiselberg \cite{hei95}; Harko et al.  \cite{harko04})    
have already dealt with thermal nucleation in hot and dense hadronic matter. 
In these studies, it has been found that the prompt formation of a critical size drop of quark matter 
via thermal activation is possible above a temperature of about $2-3$ MeV.  As a consequence, it was 
inferred that pure hadronic stars are converted to quark stars within the first seconds after their birth.  
However, these works (Horvath et al. \cite{ho92}; Horvath \cite{ho94}; Olesen \& Madsen  
\cite{ol94}; Heiselberg \cite{hei95}) reported an estimate of the thermal nucleation
based on "typical" values for the thermodynamic properties characterizing the central part of 
neutron stars.   
Quark matter nucleation during the early post-bounce stage of core collapse supernova 
has been lately examined by Sagert et al. (\cite{sage+09}) and Mintz et al. (\cite{mintz10}). 

 The main goal of the present paper is to establish whether a newborn hadronic star (proto-hadronic star) 
could survive the early stages of its evolution without ``decaying" to a quark star.  
To this aim, we calculated the thermal nucleation rate of quark matter in hot ($T \neq 0$) 
$\beta$-stable hadronic matter  using a finite temperature EOS for hadronic and quark matter.  
In addition, we calculated the quantum nucleation rate at finite temperature,   
and compared the thermal and quantum nucleation time at different temperatures and pressures  
characterizing the central conditions of metastable proto-hadronic compact stars.     
We computed the crossover temperature above which thermal nucleation dominates  the 
finite temperature quantum nucleation mechanism.   
An exploratory study of this issue has been recently reported (Bombaci et al. \cite{blppv09})   
for  the case of neutrino-free hadronic matter.   
Here, we extend our analysis and our numerical investigation to the case of hadronic matter 
with trapped neutrinos,  and we make a systematic  study of the finite temperature nucleation 
process on the value of the Bag constant in the EOS for the quark phase. 
Finally, we discuss the relevant consequences of quark matter nucleation for the evolution 
of proto-neutron stars.

\section{Equation of state}

Within a fundamental approach, the EOS for neutron star matter should  be derived by solving 
numerically the equations of QCD on a space-time lattice. 
Lattice calculations at zero baryon chemical potential (zero baryon density)  suggest that at high 
temperature and for physical values of the quark masses, the transition to quark gluon plasma 
is a ``crossover''  (Fodor \& Katz \cite{fk04}; Karsch \cite{karsch05}) rather than a phase 
transition, which would be signaled by singularities in the thermodynamic observables.   
Unfortunately, present lattice QCD calculations at finite density (baryon chemical potential) 
are plagued with the notorius ``sign problem'', which makes them unrealizable by all presently 
known lattice methods (see {\it e.g.} Lombardo (\cite{lomb07}) and references therein).     
Thus, to explore the QCD phase diagram at low temperature T and high baryon chemical potential 
$\mu$, it is necessary to invoke some approximations in QCD or to apply some QCD effective model.   
In this region of the T-$\mu$ plane, many QCD inspired models suggest the deconfinement transition 
to be a first-order phase transition (Hsu \& Schwetz \cite{hs98}, Fodor \& Katz \cite{fk04}). 
In this domain of the QCD phase diagram, many possible color superconducting phases of quark matter 
are expected (see {\it e.g.} Casalbuoni \& Nardulli \cite{CN04};  Alford et al. \cite{alf+08} and 
references therein). The effects of these color superconducting phases on the 
properties of proto-neutron stars have been investigated in (Aguilera et al. \cite{ABG04}; 
Sandin \& Blaschke \cite{SB07}; Gu et al. \cite{Gu08}; Lugones et al. \cite{lugones10} ).  

Very recently, a new phase of QCD, named quarkyonic phase, has been predicted 
(McLerran \& Pisarski \cite{McP07};  Hidaka et al. \cite{HMcP08}).   
This matter phase is characterized by chiral symmetry and confinement. 
  
Here, we have adopted a more traditional view, assuming a single first-order 
phase transition between the confined (hadronic) and deconfined phase of dense matter, 
and we  used rather common models for describing them. 

For the hadronic phase we used  the nonlinear Walecka model (NLWM) 
(Walecka \cite{wal74}, Serot \& Walecka \cite{sw86}).  This model is based on a relativistic 
Lagrangian of baryons interacting via the exchange of various mesons. 
We considered a version of the NLWM in which the baryon octet particles  
($n$, $p$, $\Lambda$, $\Sigma^{-}$, $\Sigma^{0}$, $\Sigma^{+}$, $\Xi^{-}$, $\Xi^{0}$) 
interact  via the exchange  of the scalar $\sigma$, the vector-isoscalar $\omega_\mu$ and 
the vector-isovector $\vec \rho_\mu$ meson fields (see {\it e.g.} Prakash et al. \cite{prak97}, 
Glendenning \cite{glen00}, Menezes \& Provid\^encia \cite{mp03}). 
   
The Lagrangian density of the model reads as   
 
 \begin{equation}
 {\cal L}={\cal L}_{hadrons}+{\cal L}_{leptons}
 \end{equation}
where the hadronic contribution is
  \begin{equation}
 {\cal L}_{hadrons}={\cal L}_{baryons}+{\cal L}_{mesons}
 \end{equation}
with
 \begin{equation}
 {\cal L}_{baryons}=\sum_{\mbox{baryons}} \bar \psi\left[\gamma^\mu D_\mu -M^*_B\right]\psi,
 \end{equation} 
where
 \begin{equation}
 D_\mu=i\partial_{\mu}
 -g_{\omega B} \omega_{\mu}-{g_{\rho B}} \vec{t_B} \cdot \vec{\rho}_\mu, 
 \end{equation}
and 
 $M^*_B=M_B-g_{\sigma B} \sigma.$ The quantity $\vec{t_B}$ designates the isospin of
 baryon $B$. 
The mesonic contribution reads as  
 \begin{equation}
 {\cal L}_{mesons}={\cal L}_{\sigma}+{\cal L}_{\omega}+ {\cal L}_{\rho},
 \end{equation}
 with
 \begin{equation}
     {\cal L}_\sigma=\frac{1}{2}(\partial_{\mu}\sigma\partial^{\mu}\sigma
     -m_{\sigma}^2 \sigma^2)+ \frac{1}{3!} \kappa \sigma^3+ \frac{1}{4!} \lambda
     \sigma^4,
 \end{equation}
 \begin{equation}
     {\cal L}_{\omega}=-\frac{1}{4}\Omega_{\mu\nu}\Omega^{\mu\nu}+\frac{1}{2}
     m_{\omega}^2 \omega_{\mu}\omega^{\mu}, \qquad \Omega_{\mu\nu}=\partial_{\mu}\omega_{\nu}-\partial_{\nu}\omega_{\mu},
 \end{equation}
 \begin{equation}
     {\cal L}_{\rho}=
    { -\frac{1}{4}\vec B_{\mu\nu}\cdot\vec B^{\mu\nu}}+\frac{1}{2}
     m_\rho^2 \vec \rho_{\mu}\cdot \vec \rho^{\mu}, \quad \vec B_{\mu\nu}=\partial_{\mu}\vec \rho_{\nu}-\partial_{\nu} \vec \rho_{\mu}
       - g_\rho (\vec \rho_\mu \times \vec \rho_\nu)
 \end{equation}
 For the lepton contribution we take
 \begin{equation}
 {\cal L}_{leptons}=\sum_{\mbox{leptons}} \bar \psi_l \left(i \gamma_\mu \partial^{\mu}-
 m_l\right)\psi_l,
 \end{equation}
where the sum is over electrons, muons and neutrinos for matter with trapped neutrinos.  
 In uniform matter, we get for the baryon Fermi energy
 $
       \epsilon_{FB}=g_{\omega B} \omega_0+ g_{\rho B} t_{3B} \rho_{03} + \sqrt{k_{FB}^2+{M^*_B}^2},
 $
 with the baryon effective mass
    $M^*_B=M-g_{\sigma B}\sigma.$

In the present work we  used one of the parametrizations of the NLWM given by  
Glendenning \& Moszkowski (\cite{gm91}) (hereafter the GM1 equation of state).      
The parameters of the GM1 EOS are fitted to the saturation properties of symmetric nuclear matter:   
binding energy per nucleon $B/A = -16.3$~MeV, saturation density $\rho_0 = 0.153$~fm$^{-3}$,  
incompressibility $K = 300$~MeV, nucleon effective mass $M^* = 0.7$, and to the nuclear  
symmetry energy at saturation  density $a_{sym} = 32.5$~MeV. 

Including  hyperons involves new couplings, i.e., the hyperon-nucleon couplings:
 $g_{\sigma B}=x_{\sigma B}~ g_{\sigma},~~g_{\omega B}=x_{\omega B}~ g_{\omega},~~g_{\rho B}=x_{\rho B}~ g_{\rho}$.
 For nucleons we take
  $x_{\sigma B}$, $x_{\omega B}$, $x_{\rho B} = 1$
 and for hyperons we will consider the couplings proposed by Glendenning and Moszkowski (\cite{gm91}). 
They consider the binding energy of the $\Lambda$ in nuclear  matter, $B_\Lambda$, 
 \begin{equation}
 \left(\frac{B_\Lambda}{A}\right)=-28 \mbox{ MeV}= x_{\omega} \, g_{\omega}\, \omega_0-x_{\sigma}\, g_{\sigma} \sigma
 \end{equation} 
to establish a relation between $x_{\sigma}$ and $x_{\omega}$.  
Moreover, known neutron star masses restrict $x_{\sigma}$ to the range $0.6-0.8$. We take 
$x_{\rho}=x_{\sigma}$ and consider $x_{\sigma}=0.6$.

For the quark phase,  we  adopted a phenomenological EOS (Farhi \& Jaffe \cite{fj84}) that is 
based on the MIT bag model for hadrons and considered different possible values 
for the bag constant B. The remaining model parameters have been fixed to: 
$m_u = m_d =0$, $m_s = 150$ MeV for the masses of the {\it up}, {\it down}, and {\it strange} quark,  respectively, and  $\alpha_s = 0$ for the QCD structure constant. 

The two models for the EOS have been generalized to the case of finite temperature 
(see {\it e.g.} Menezes \& Provid\^encia \cite{mp03}).

\section{Phase equilibrium}
 
For a first-order phase transition, the conditions for phase equilibrium  are given by the Gibbs' phase rule
\begin{equation}
T_H  = T_Q \equiv T   \, ,~~~~~~~~
P_H = P_Q \equiv P_0 \, ,   
\label{eq:eq1a}
\end{equation}
\begin{equation}
\mu_H(T, P_0)  =  \mu_Q(T, P_0) \, , 
\label{eq:eq1b}
\end{equation}
where 
\begin{equation}
  \mu_H = \frac{\varepsilon_H + P_H - s_H T}{n_H}  \, , ~~
  \mu_Q = \frac{\varepsilon_Q + P_Q - s_Q T}{n_Q}  
\label{eq:eq2}
\end {equation} 
are the Gibbs' energies per baryon (average chemical potentials) for the hadron and 
quark phase respectively, and 
$\varepsilon_H$ ($\varepsilon_Q$),  $P_H$ ($P_Q$), $s_H$ ($s_Q$)  and $n_{H}$  ($ n_{Q}$)
respectively denote the total ({\it i.e.,}  including leptonic contributions) energy 
density, total pressure, total entropy density,  and baryon number density  for the hadron (quark)  
phase.  
Above the ``transition point" ($P_0$), the hadronic phase is metastable,  
and the stable quark phase will appear  as a result  of a nucleation process. 

Small localized  fluctuations in the state variables of the metastable hadronic phase  
will give rise to virtual drops of the stable quark phase. These fluctuations  are characterized 
by a time scale  $\nu_0^{-1} \sim 10^{-23}$ s.  It  is set by the strong 
interactions (responsible for the deconfinement phase transition), 
and it is many orders of magnitude shorter than the typical time scale for the weak interactions.  
Quark flavor must therefore be conserved during the deconfinement transition.   
We refer to this form of deconfined matter,  in which the flavor content is equal to that of 
the $\beta$-stable hadronic system at the same pressure and temperature, as the Q*-phase. 
Soon after a critical size drop of quark matter is formed, the weak interactions  
will have enough time to act, changing the quark flavor fraction of the deconfined droplet to lower 
its energy, and a droplet of $\beta$-stable quark matter is formed (hereafter the Q-phase).

This first seed of quark matter will trigger the conversion (Olinto \cite{oli87}; 
Heiselberg et al. \cite{hbp91}; Bombaci \& Datta \cite{grb}; Drago et al. \cite{DLP07})   
of the pure hadronic star to a hybrid star or to a strange star.  
Thus, pure hadronic stars with values of the central pressure higher than  $P_0$ are metastable to 
the decay (conversion) to hybrid stars or to strange stars (Berezhiani et al. \cite{be02}; \cite{be03}; 
Bombaci et al. \cite{bo04}; Drago et al. \cite{drago04}; Lugones \& Bombaci \cite{lug05}; 
Bombaci et al. \cite{blv07}). 
The mean lifetime of the metastable stellar configuration is related to the time needed to 
nucleate the first drop of quark matter in the stellar center, and it depends dramatically 
on the value of the stellar central pressure.  

%%%%%%%%%%%%%%%%%%%% FIG. 1 %%%%%%%%%%%%%%%%%%%%%%%%% 
%                                                One column figure
%----------------------------------------------------------- S_vib
\begin{figure}
\centering
\vspace*{18pt}
\includegraphics[width=8cm]{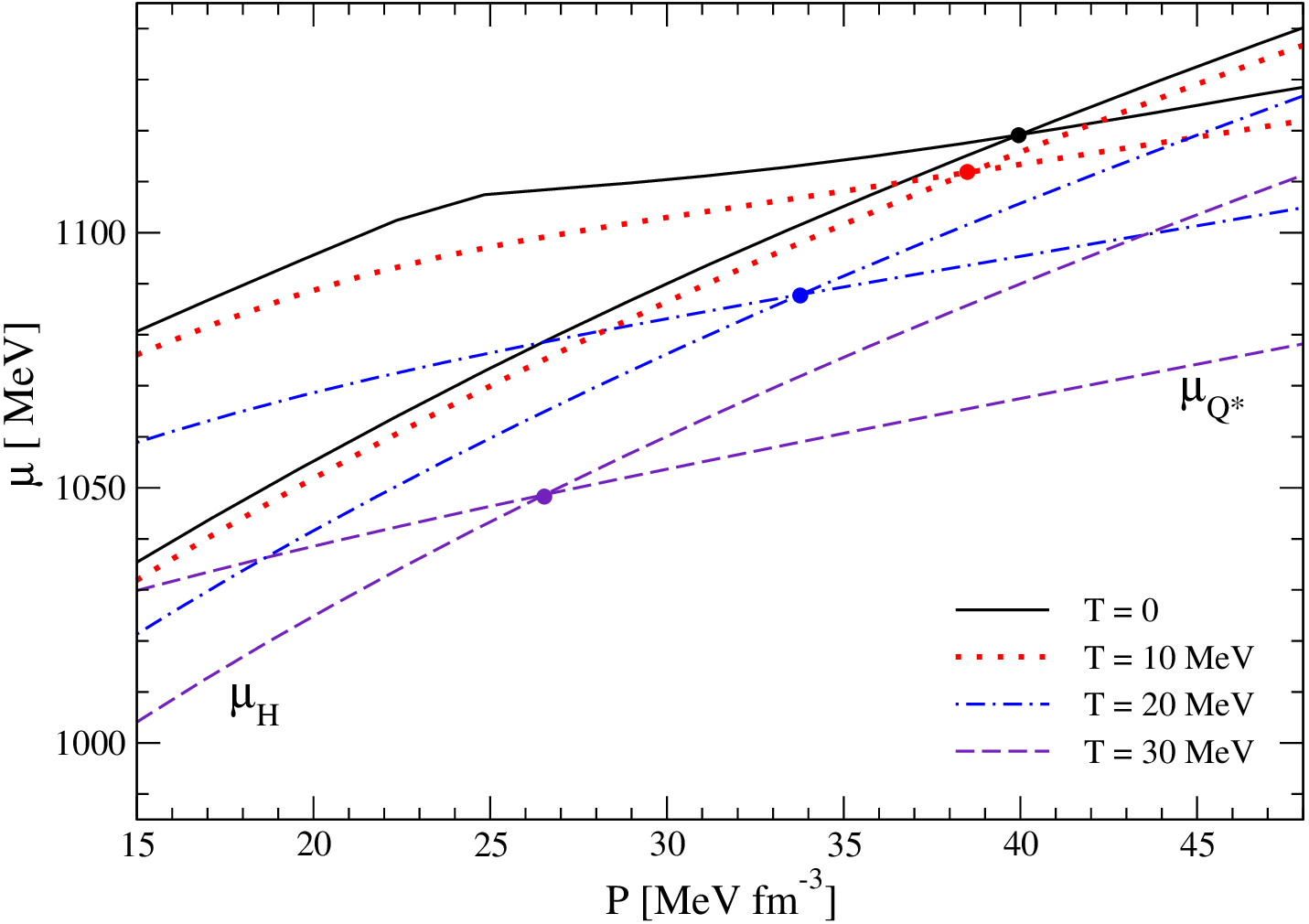}  %gibbs_GM1_B85.eps
\caption{(Color online.) Gibbs energy per baryon, in neutrino-free matter, as a function of pressure 
for the hadronic and quark phases at different temperatures. 
Lines with the steeper slope refer to the hadronic phase. 
Full dots indicate the transition pressure $P_0$ for each temperature. 
GM1 EOS for the hadronic phase and MIT bag model EOS with B = 85 MeV/fm$^3$ for the quark phase.} 
\label{Fig1}
\end{figure}
%______________________________________________________________

In Figs. 1 and  2,  we plot the Gibbs' energies per baryon for the hadron-phase and the Q*-phase 
in neutrino-free matter, at different temperatures (T = 0, 10, 20, 30 MeV) and for two different 
values of the bag constant B = 85 MeV/fm$^3$ (Fig. 1) and  B = 100 MeV/fm$^3$ (Fig. 2).  
Lines with the steeper slope refer to the hadron phase.  As we see, the transition pressure $P_0$ 
(indicated by a full dot) decreases when the hadronic matter temperature is increased.   

  The phase equilibrium curve  $P_0(T)$ for the hadron-quark phase transition 
(in the case B = 85 MeV/fm$^3$) is shown in Fig. \ref{Fig3} for neutrino-free matter and matter 
with trapped neutrinos. The region of the  $P_0$--$T$ plane above each curve represents the 
deconfined Q*-phase.  As expected (Prakash et al. \cite{prak97}; Lugones \& Benvenuto \cite{lb98}; 
Vida\~na et al. \cite{vbp05}; Lugones et a. \cite{lugones09}) neutrino trapping in $\beta$-stable 
hadronic matter inhibits the quark deconfinement phase transition, thus the global effect of 
neutrino-trapping is to produce a shift of the phase equilibrium curve toward higher values of 
the pressure in the  $P_0$--$T$ plane.   

As  is well known, for a first-order phase transition the derivative $dP_0/dT$ is related to the 
specific  ({\it i.e.} per baryon) latent heat ${\cal Q}$  of the phase 
transition by the  Clapeyron-Clausius equation   
\begin{equation}
            \frac{dP_0}{dT} =  - \frac{ n_H n_{Q^*}}{n_{Q^*} - n_H } \frac{{\cal Q}}{T } 
\label{eq:eq10}
\end {equation}
\begin{equation}
           {\cal Q}  = \tilde W_{Q^*} - \tilde W_H = T(\tilde {S}_{Q^*} - \tilde {S}_H) 
\label{eq:eq11}
\end {equation}
where $\tilde W_H$ ($\tilde W_{Q^*}$)  and $\tilde {S}_H$ ($\tilde {S}_{Q^*}$) denote the  enthalpy  
per baryon and entropy per baryon for the hadron (quark) phase, respectively.   
The specific latent heat ${\cal Q}$ and the phase numbers densities  $n_H$ and  $n_{Q^*}$ 
at phase equilibrium are reported in Tables 1 and 2 for the case of neutrino-free matter    
and for two different values of the bag constant B = 85 MeV/fm$^3$ (Table 1) and  
B = 100 MeV/fm$^3$ (Table 2).  
As expected for a first-order phase transition, one has a discontinuity jump in the phase 
number densities:  in our particular case  $n_{Q^*}(T,P_0) > n_H(T,P_0)$. 
This result, together with the positive value of ${\cal Q}$ ({\it i.e.} the deconfinement phase 
transition absorbs heat), tells us (see Eq.\ (\ref{eq:eq10})) that the phase transition temperature   
(\ref{eq:eq1a}) decreases with pressure (as in the melting of ice).    

To explore the effect of neutrino trapping on the phase equilibrium properties of the system, in 
Table 3 we report the quantities  ${\cal Q}$, $n_H$, $n_{Q^*}$ and $P_0$ at different temperatures, 
for a bag constant B = 85 MeV/fm$^3$ (compare with the results in Table 1).  
As we see, the phase number  densities  $n_H$ and  $n_{Q^*}$ at phase equilibrium are shifted 
to higher values, and the specific latent heat ${\cal Q}$  is increased with respect to the neutrino-free 
matter case.   

We also made a systematic calculations of the phase equilibrium properties varying the value of the bag constant in the EOS of the Q*-phase. These results are summarized in Fig.~4 where we plot the phase equilibrium pressure $P_0$ between the hadron phase and the Q* phase at zero temperature as a 
function of $B$ for neutrino-free matter and  neutrino-trapped matter.  As expected, a higher value of 
$B$ makes the deconfinemnt transition more difficult, increasing the value of the transition pressure.  
%%%%%%%%%%%%%%%%%%%% FIG. 2 %%%%%%%%%%%%%%%%%%%%%%%%% 
%                                                One column figure
%----------------------------------------------------------- S_vib
\begin{figure}
\centering
\vspace*{18pt}
\includegraphics[width=8cm] {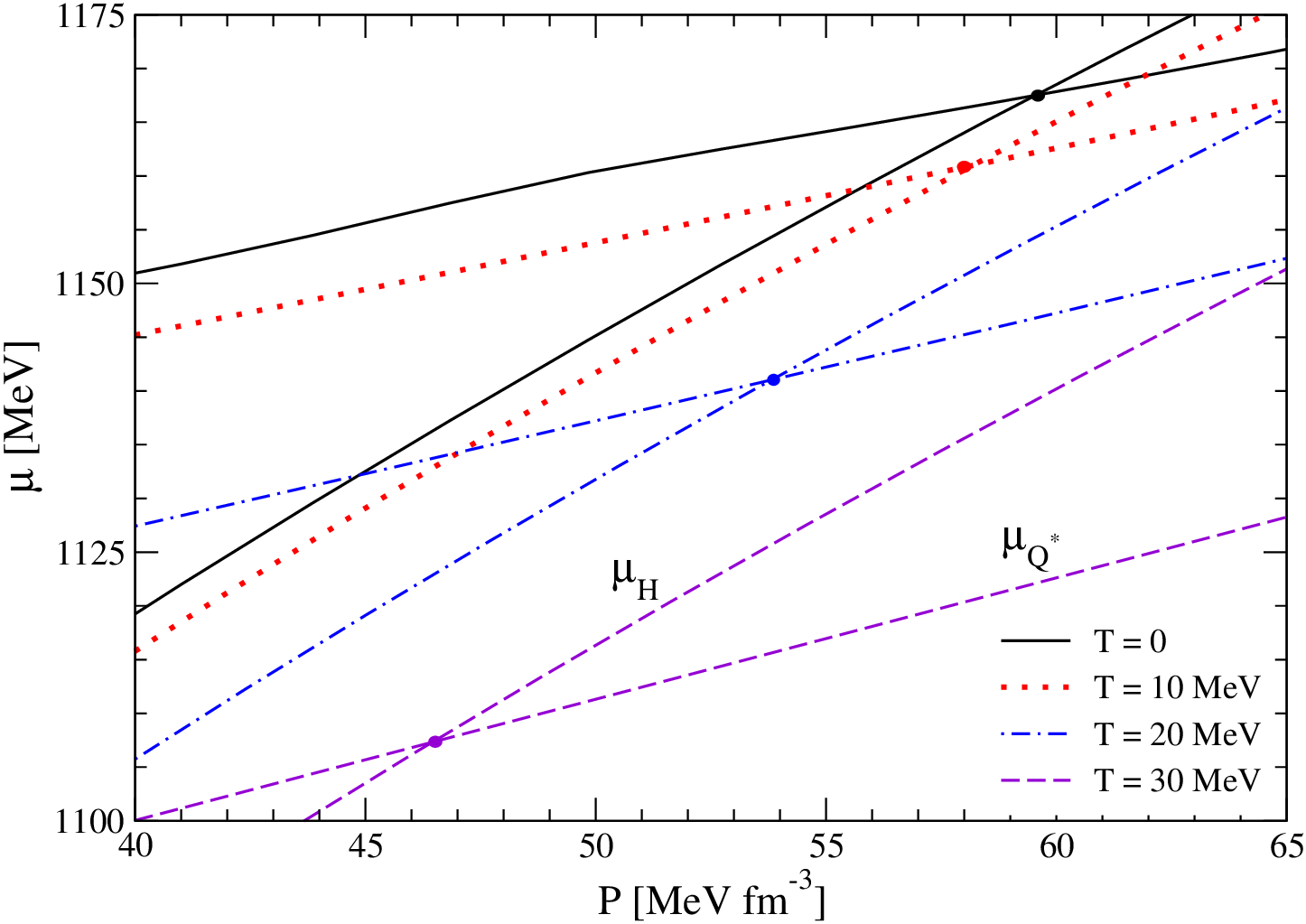}       %gibbs_GM1_B100.eps
\caption{Same as Fig. 1, but with a bag constant B = 100 MeV/fm$^3$. (Color online)} 
\end{figure}
%______________________________________________________________

%_____________________________________________________________
%                                             Simple A&A Table
%_____________________________________________________________
%%%%%%%%%   TABLE 1   %%%%%%%%%%%%%%%%%%%%%%%
\begin{table}
\caption{The specific latent heat ${\cal Q}$ and the phase number  densities  $n_H$ and  $n_{Q^*}$ 
at phase equilibrium. GM1 EOS for the hadronic phase, MIT bag model for the quark phase 
with $B = 85$~MeV/fm$^3$. Results for neutrino-free matter.}       % title of Table
\label{table:latent1}      % is used to refer this table in the text
\centering                          % used for centering table
\begin{tabular}{c c c c c}     % centered columns (5 columns)
\hline\hline                 % inserts double horizontal lines
 $T$~~~ & ${\cal Q}$~~~~  & $n_{Q^*}$~~~~  & $n_{H}$~~~~ & $P_0$\\
 MeV~~~ & MeV~~~~ & fm$^{-3}$~~~~ & fm$^{-3}$~~~~ & MeV/fm$^3$ \\ % table heading 
\hline                        % inserts single horizontal line
  0~~~     &    0.00~~~~ &     0.453~~~~&  0.366~~~~  &     39.95 \\   % inserting body of the table
  5~~~     &    0.56~~~~ &     0.451~~~~&  0.364~~~~  &     39.74 \\  
10~~~     &    2.40~~~~ &     0.447~~~~&  0.358~~~~  &     38.58 \\
15~~~     &    5.71~~~~ &     0.439~~~~&  0.348~~~~  &     36.55 \\
20~~~     &  10.60~~~~ &     0.428~~~~&  0.334~~~~  &     33.77 \\
25~~~     &  17.17~~~~ &     0.414~~~~&  0.316~~~~  &     30.36 \\
30~~~     &  25.44~~~~ &     0.398~~~~&  0.294~~~~  &    26.53 \\
\hline                                   %inserts single line
\end{tabular}
\end{table}
%
%_____________________________________________________________

%_____________________________________________________________

%%%%%   TABLE 2   %%%%%%%%%%%%%%%%%%%%%%%
\begin{table}
\caption{Same as Table 1, but with $B = 100$~MeV/fm$^3$. }       % title of Table
\label{table:latent2}      % is used to refer this table in the text
\centering                          % used for centering table
\begin{tabular}{c c c c c}     % centered columns (5 columns)
\hline\hline                 % inserts double horizontal lines
 $T$~~~ & ${\cal Q}$~~~~  & $n_{Q^*}$~~~~  & $n_{H}$~~~~ & $P_0$\\
 MeV~~~ & MeV~~~~ & fm$^{-3}$~~~~ & fm$^{-3}$~~~~ & MeV/fm$^3$ \\ % table heading 
\hline                        % inserts single horizontal line
  0~~~     &    0.00~~~~ &     0.557~~~~&  0.447~~~~  &     59.53 \\   % inserting body of the table
  5~~~     &    0.44~~~~ &     0.556~~~~&  0.445~~~~  &     59.22 \\  
10~~~     &    1.98~~~~ &     0.552~~~~&  0.441~~~~  &     58.18 \\
15~~~     &    4.91~~~~ &     0.544~~~~&  0.432~~~~  &     56.39 \\
20~~~     &    9.35~~~~ &     0.533~~~~&  0.421~~~~  &     53.81 \\
25~~~     &  15.49~~~~ &     0.519~~~~&  0.401~~~~  &     50.47 \\
30~~~     &  23.51~~~~ &     0.503~~~~&  0.386~~~~  &     46.44\\
\hline                                   %inserts single line
\end{tabular}
\end{table}
%
%_____________________________________________________________

%%%%%   TABLE 3  %%%%%%%%%%%%%%%%%%%%%%%
\begin{table}
\caption{Same as Table 1, but with trapped neutrinos.}       % title of Table
\label{table:latent3}      % is used to refer this table in the text
\centering                          % used for centering table
\begin{tabular}{c c c c c}     % centered columns (5 columns)
\hline\hline                 % inserts double horizontal lines
 $T$~~~ & ${\cal Q}$~~~~  & $n_{Q^*}$~~~~  & $n_{H}$~~~~ & $P_0$\\
 MeV~~~ & MeV~~~~ & fm$^{-3}$~~~~ & fm$^{-3}$~~~~ & MeV/fm$^3$ \\ % table heading 
\hline                        % inserts single horizontal line
  0~~~     &    0.00~~~~ &     0.603~~~~&  0.516~~~~  &    113.77 \\   % inserting body of the table
  5~~~     &    0.65~~~~ &     0.601~~~~&  0.514~~~~  &    113.11 \\  
10~~~     &    2.87~~~~ &     0.594~~~~&  0.509~~~~  &    110.69 \\
15~~~     &    6.78~~~~ &     0.580~~~~&  0.499~~~~  &    106.17 \\
20~~~     &  12.65~~~~ &     0.560~~~~&  0.483~~~~  &      99.18 \\
25~~~     &  20.21~~~~ &     0.534~~~~&  0.462~~~~  &      90.12 \\
30~~~     &  29.88~~~~ &     0.502~~~~&  0.434~~~~  &      78.65 \\
\hline                                   %inserts single line
\end{tabular}
\end{table}
%
%_____________________________________________________________

%%%%%%%%%%%%%%%%%%%% FIG. 3 %%%%%%%%%%%%%%%%%%%%%%%%% 
%                                                One column figure
%----------------------------------------------------------- S_vib
\begin{figure}
\centering
\vspace*{18pt}
\includegraphics[width=8cm]{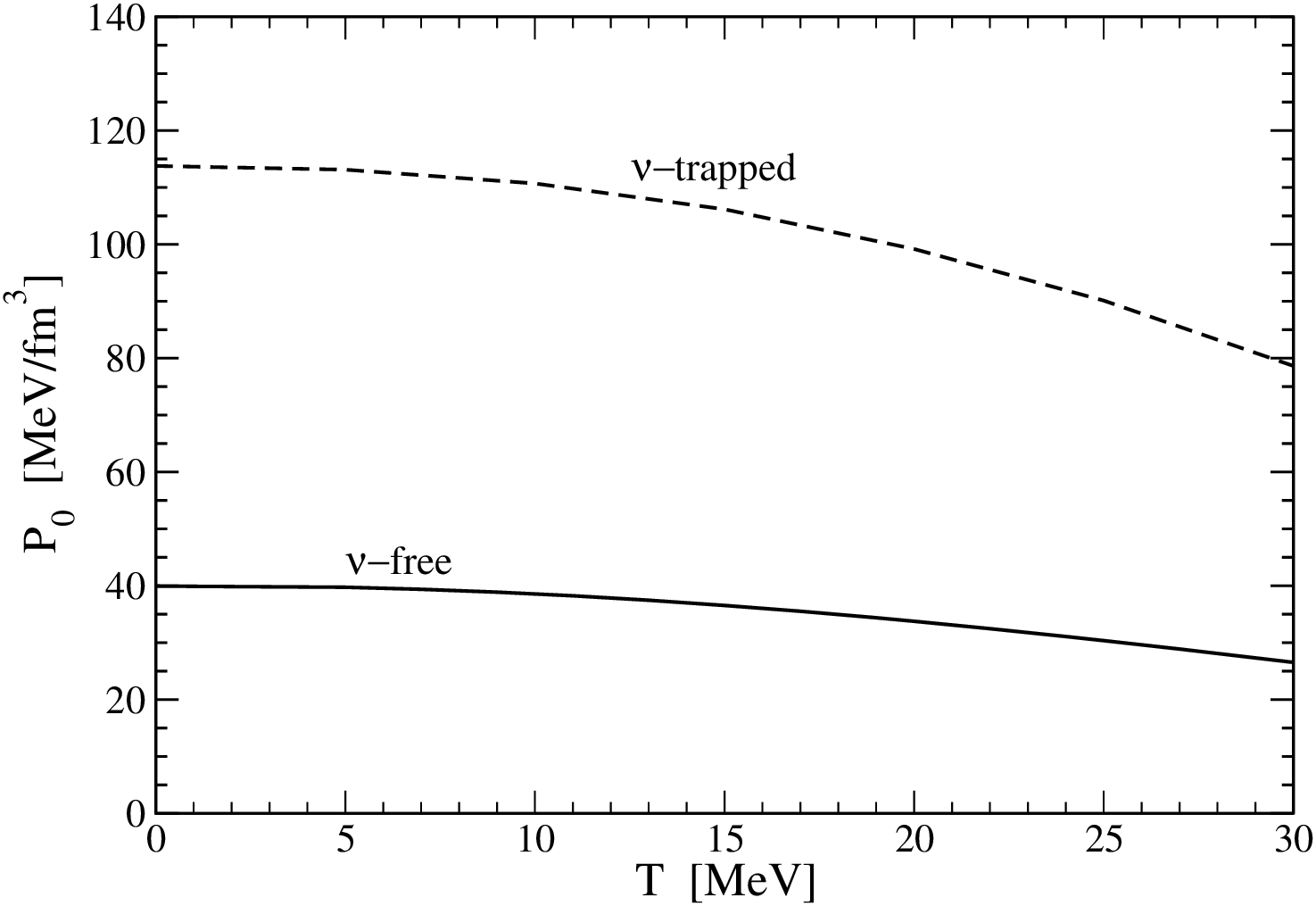}   %p0-T_GM1_B85.eps
\caption{Phase equilibrium curve for the hadron to Q*-matter phase transition.
The continuous curve is relative to neutrino-free matter, the dashed curve to matter with 
trapped  neutrinos. EOS: GM1 plus MIT Bag model with B = 85 MeV/fm$^3$.}  
\label{Fig3}
\end{figure}
%%%%%%%%%%%%%%%%%%%%%%%%%%%%%%%%%%%%%%%%%%%%%%%%%

%%%%%%%%%%%%%%%%%%%% FIG. 4 %%%%%%%%%%%%%%%%%%%%%%%%% 
%                                                One column figure
%----------------------------------------------------------- S_vib
\begin{figure}
\centering
\vspace*{18pt}
\includegraphics[width=8cm]{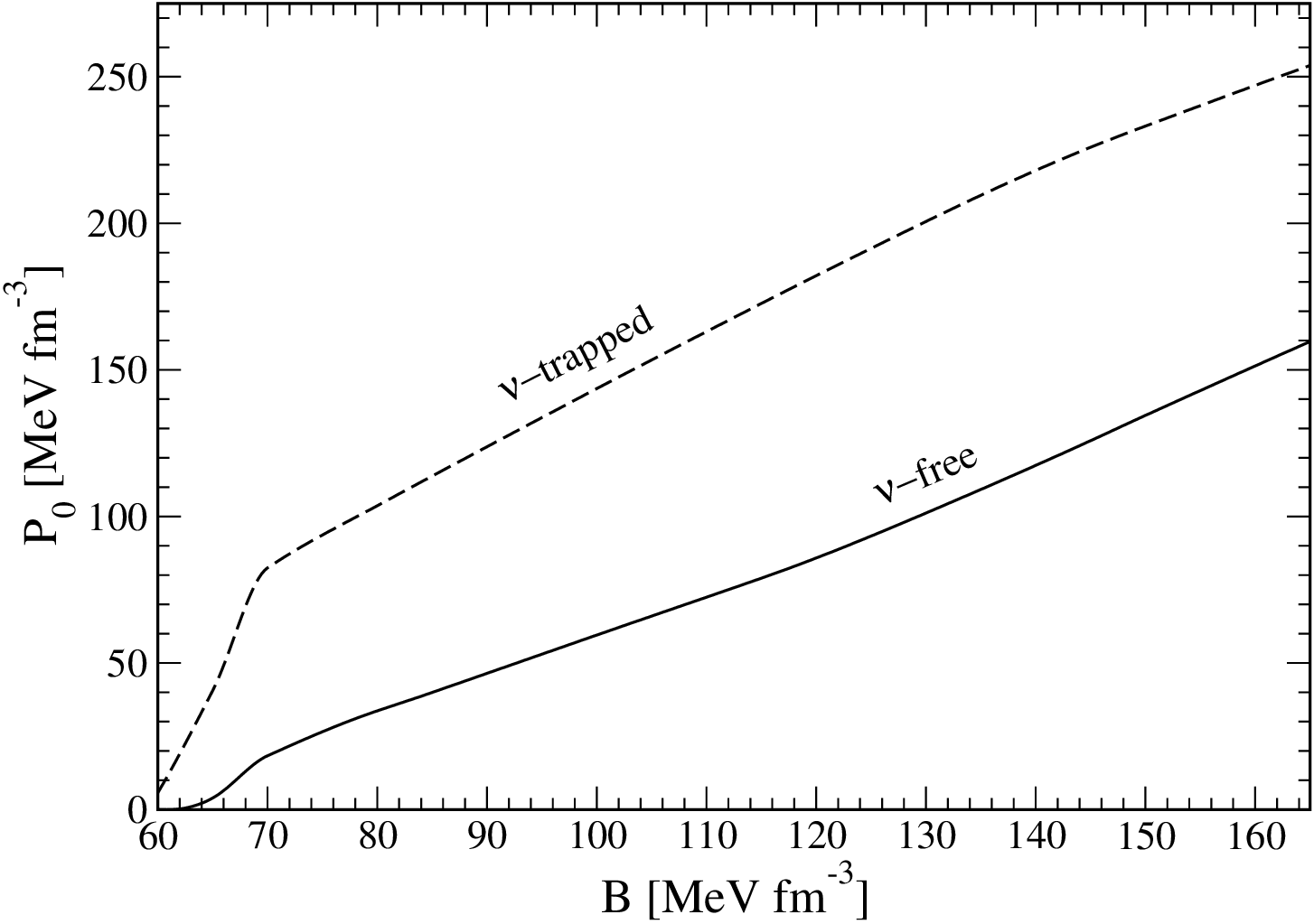}      %{p0_B.eps}
\caption{Phase equilibrium pressure $P_0$  between the  hadron phase and the Q* phase 
at T = 0, as a function of the bag constant $B$.    
The continuous curve refers to neutrino-free matter, and the dashed curve refers to 
neutrino-trapped matter. }  
\label{Fig4}
\end{figure}
%______________________________________________________________

%%%%%%%%%%%%%%%%%%%%%%%%%%%%%%%%%%%%%%%%%%%%%%%%%

\section{Quantum and thermal nucleation rates} 

Above the phase equilibrium pressure $P_0$ the  $\beta$-stable hadron phase is metastable,  
and the formation of the stable (with respect to the strong interactions) Q*-phase will occur via  
a nucleation process. The main effect of finite temperature on the quantum nucleation mechanism 
of quark matter is to modify the energy barrier separating the quark phase from the metastable 
hadronic phase. This energy barrier, which represents the difference in the free energy of the system 
with and without a  Q*-matter droplet, can be written as 
\begin{equation}
  U({\cal R}, T) = \frac{4}{3}\pi n_{Q^*}(\mu_{Q^*} - \mu_H){\cal R}^3 + 4\pi \sigma {\cal R}^2
\label{eq:potential}
\end{equation}
where ${\cal R}$ is the radius of the droplet (supposed to be spherical), and $\sigma$ is 
the surface tension for the surface separating the hadron from the Q*-phase. 
The energy barrier has a maximum at the critical radius 
${\cal R}_c = 2 \sigma /[n_{Q^*}(\mu_H - \mu_{Q^*})]$.    
We neglected the term associated with the curvature energy, and also the terms connected with 
the electrostatic energy, since they are known to only introduce small corrections 
(Iida \& Sato \cite{iida98}; Bombaci et al. \cite{bo04}).   
The value of the surface tension $\sigma$ for the interface separating the quark and hadron phase 
is poorly known, and typically values used in the literature range within $10-50$ MeV fm$^{-2}$ 
(Heiselberg et al. \cite{hei93}; Iida \& Sato \cite{iida98}). 
We assume $\sigma$ to be temperature independent and we take $\sigma = 30$ MeV fm$^{-2}$.    

The quantum nucleation time $\tau_q$ can be straightforwardly evaluated within a semi-classical 
approach (Lifshitz \& Kagan \cite{lk72}; Iida \& Sato \cite{iida97}; \cite{iida98}). 
First one computes the ground state energy $E_0$ in the WKB approximation and 
the oscillation frequency $\nu_0$ of the drop in the potential well $U({\cal R},T)$.   
Then, the probability of tunneling is given by 
\begin{equation}
  p_0=exp\left[-\frac{A(E_0)}{\hbar}\right]
\label{eq:prob}
\end{equation}
where $A(E)$ is the action under the potential barrier, which in a relativistic framework 
reads as (Iida \& Sato \cite{iida98})   
\begin{equation}
 A(E)=\frac{2}{c}\int_{{\cal R}_-}^{{\cal R}_+}\sqrt{[2m({\cal R})c^2 +E-U({\cal R})][U({\cal R})-E]}   
\label{eq:action}
\end{equation} 
with ${\cal R}_\pm$ the classical turning points and $m({\cal R})$ the droplet effective mass. 
The quantum nucleation time is then equal to
\begin{equation}
  \tau_q  = (\nu_0 p_0 N_c)^{-1} \ , 
\label{eq:time}
\end{equation} 
with $N_c \sim 10^{48}$ the number of nucleation centers expected in the innermost part 
($r \leq R_{nuc} \sim100$ m) of the hadronic star, where the pressure and temperature 
can be considered constant and equal to their central values.    

The thermal nucleation rate can be written (Langer \& Turski \cite{LanTur73}) as 
\begin{equation}
    I =\frac{\kappa}{2 \pi} \Omega_0 \exp (- U({\cal R}_c, T) /T)
\label{eq:therm_rate}
\end{equation}
where $\kappa$ is the so-called dynamical prefactor, which is related to the growth rate of the 
drop radius $\cal R$ near the critical radius (${\cal R}_c$);  $\Omega_0$ is the so-called statistical 
prefactor, which measures the phase-space volume of the saddle-point region around ${\cal R}_c$;  
and $U({\cal R}_c, T)$ is the activation energy, {\it i.e.} the change in the free energy of the system 
required to activate the formation of a critical size droplet. 
The Langer theory (Langer \cite{lang68}; \cite{lang69}; Langer \& Turski \cite{LanTur73};
Turski \& Langer \cite{TurLan80}) of homogeneous nucleation has been 
extended in Csernai \& Kapusta (\cite{CseKap92}) and  Venugopalan \& Vischer (\cite{VenVis94})   
to the case of first-order phase transitions occurring in relativistic systems, as in the case of the 
quark deconfinement transition. 
The statistical prefactor  can be written (Csernai \& Kapusta \cite{CseKap92})  as 
\begin{equation}
   \Omega_0 = 
\frac{2}{3\sqrt{3}}  \Big(\frac{\sigma}{T}\Big)^{3/2}  \Big(\frac{\cal R}{\xi_Q}\Big)^4
\label{eq:omega}
\end{equation}
where $\xi_Q$ is the quark correlation length, which gives a measure of the thickness of the 
interface layer  between the two phases (the droplet ``surface thickness"). 
In the present calculation, we take  $\xi_Q = 0.7$~fm according to the estimate given in 
Csernai \& Kapusta (\cite{CseKap92}) and  Heiselberg (\cite{hei95}).  
For the dynamical  prefactor, we have used a general expression, which has been derived by 
Venugopalan \& Vischer (\cite{VenVis94}) 
\begin{equation}
\kappa  = \frac{2 \sigma} {{\cal R}_c^3 (\Delta w)^2} \Big [ \lambda T + 2 \Big(\frac{4}{3} \eta + \zeta \Big)\Big]  \,  ,
\label{eq:kappa}
\end{equation} 
where $\Delta w = w_{Q*} - w_H$ is the difference between the enthalpy density of the two phases, 
$\lambda$  the thermal conductivity, and $\eta$ and $\zeta$ are the shear and bulk viscosities 
respectively of hadronic matter. The nucleation prefactor used in the present work differs significantly 
from the one used in previous works (Horvath et al. \cite{ho92}; Horvath \cite{ho94}; 
Olesen \& Madsen \cite{ol94}), where based on dimensional grounds, the prefactor was taken to be 
equal to $T^4$.  

There are not many calculations of the transport properties of dense hadronic matter. 
With a few exceptions (see {\it e.g.} van Dalen \& Dieperink \cite{vDD04}; 
Chattaerjee \& Bandyopadhyay \cite{CB06}) most of them are relative to nuclear or pure neutron matter 
(Flowers \&Itho \cite{FI79}; Danielewicz \cite{Dan84}; Sedrakian et al. \cite{Sedr94}; 
Benhar \& Valli \cite{Benh07}; Shternin \& Yakovlev \cite{Yakov08}; Benhar et al. \cite{Benh10}; 
Zhang et al. \cite{ZLZ10}).   
These quantities have been calculated by Danielewicz \cite{Dan84} in the case of nuclear matter.  
According to the results of Danielewicz \cite{Dan84}, the dominant contribution to the prefactor 
$\kappa$ comes from the shear viscosity $\eta$. Therefore, we take $\lambda$ and $\zeta$ equal to zero,
and we use the following relation for the shear viscosity (Danielewicz \cite{Dan84}): 
\begin{equation}
\eta =  \frac{7.6 \times 10^{26}} {(T/{\rm MeV})^2}  \Big(\frac{n_H}{n_0}\Big)^2 ~~ 
            \frac{{\rm MeV}}{{\rm fm \, s}}  \,  ,
\label{eq:eta} 
\end{equation}                            
with $n_0 = 0.16$~fm$^{-3}$  the saturation density of normal nuclear matter. 

 The thermal nucleation time $\tau_{th}$, relative to the innermost stellar region 
($V_{nuc} = (4 \pi/3) R_{nuc}^3$) where almost constant pressure and temperature occur, can thus 
be written as   
\begin{equation}
    \tau_{th} = (V_{nuc} \, I )^{-1} . 
\end{equation}

 In Fig.~5, we represent the energy barrier for a virtual drop of the Q*-phase (with B = 85 MeV/fm$^3$)
in the neutrino-free hadronic phase as a function of the droplet radius and for different 
temperatures at a fixed  pressure $P = 57$~MeV/fm$^3$. 
 As expected, from the results plotted in Fig. 1,  the energy 
barrier $U({\cal R}, T)$ and the droplet critical radius  ${\cal R}_c$ decrease as the matter 
temperature is increased. This effect favors the Q*-phase formation and, in particular,  increases 
(decreases) the quantum nucleation rate (nucleation time $\tau_q$) with respect to the 
corresponding quantities  calculated at  $T=0$.   

%%%%%%%%%%%%%%%%%%%% FIG. 5 %%%%%%%%%%%%%%%%%%%%%%%%% 
%                                                One column figure
%----------------------------------------------------------- S_vib
\begin{figure}
\centering
\vspace*{18pt}
\includegraphics[width=8cm] {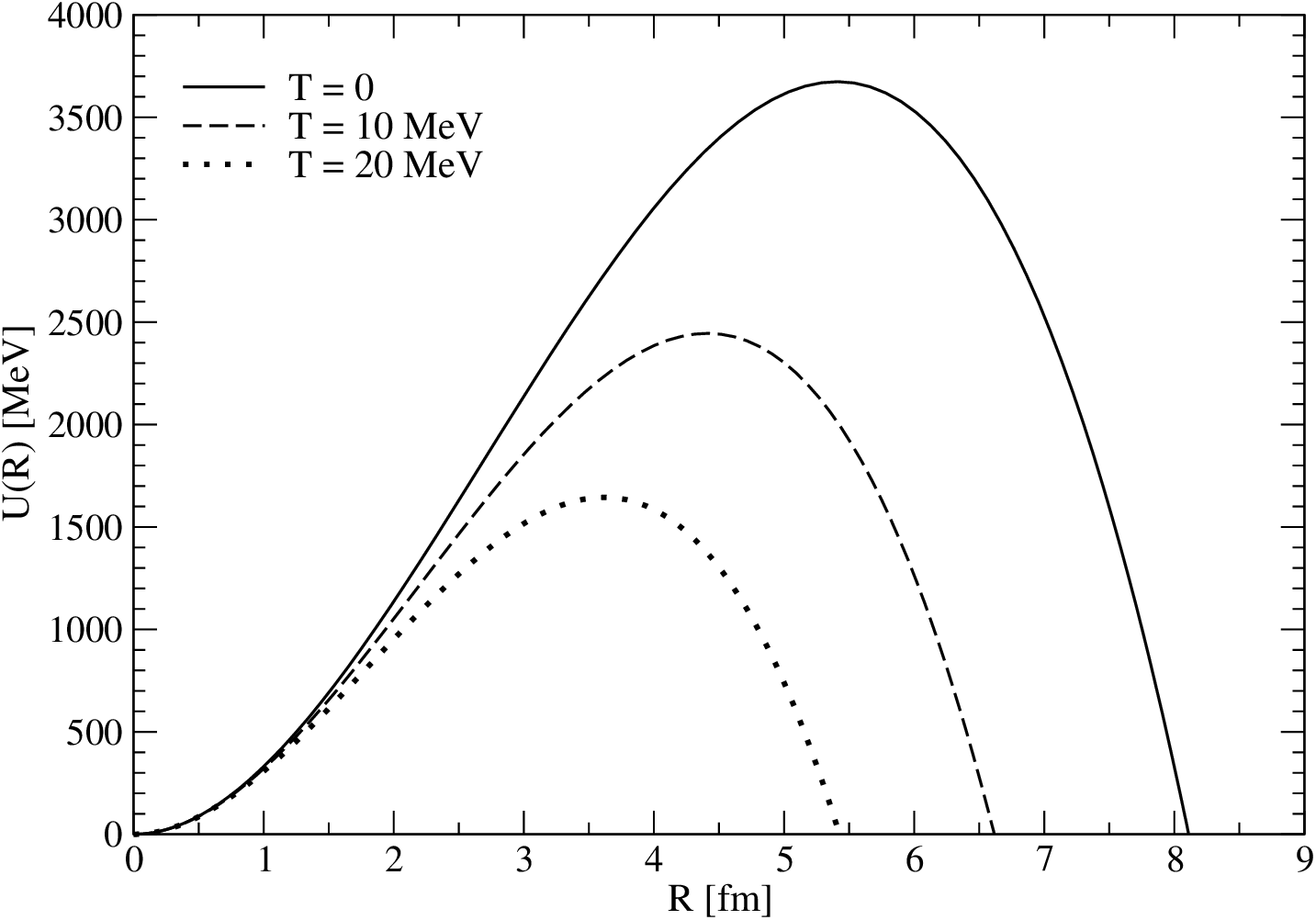}  %{barrier_T.eps}
\caption{Energy barrier for a virtual drop of the Q*-phase  in $\beta$-stable 
neutrino-free hadronic matter as a function of the droplet radius 
and for different temperatures for a fixed pressure $P = 57$~MeV/fm$^3$.  
The bag constant takes the value  B = 85 MeV/fm$^3$. } 
\label{Fig5}
\end{figure}
%______________________________________________________________
%%%%%%%%%%%%%%%%%%%%%%%%%%%%%%%%%%%%%%%%%%%%%%%%%

In Fig.~6  we plot the quantum and thermal nucleation times of the  Q*-phase 
(with B = 85 MeV/fm$^3$) in  $\beta$-stable neutrino-free hadronic matter 
as a function of temperature and at a fixed pressure $P=57$~MeV/fm$^3$. 
As expected, we find a crossover temperature $T_{co}$ above which thermal nucleation is dominant 
with respect to the quantum nucleation mechanism.  For the case reported in Fig.~6, we have 
$T_{co}=7.05$~MeV and the corresponding nucleation time is $\log_{10}(\tau/{\rm s}) = 54.4$.
The crossover temperature for different values of the pressure of $\beta$-stable hadronic matter    
is reported in Table~4 (second column), together with the nucleation time calculated at  
$T = T_{co}$  (third column).  

%%%%%%%%%%%%%%%%%%%% FIG. 6 %%%%%%%%%%%%%%%%%%%%%%%%% 
%                                                One column figure
%----------------------------------------------------------- S_vib
\begin{figure}
\centering
\vspace*{18pt}
\includegraphics[width=8cm] {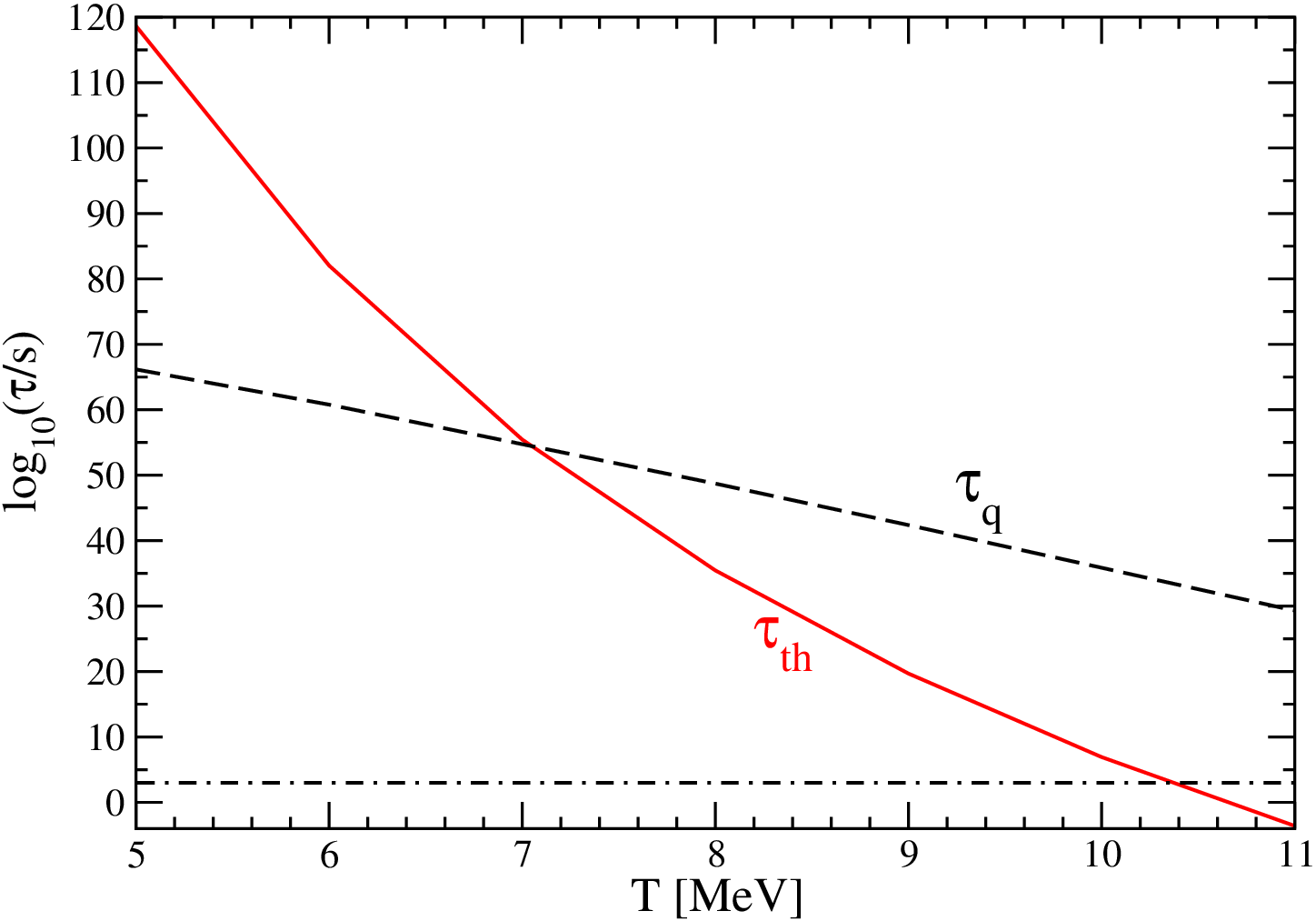}    %{time_57.eps}
\caption{(Color online.) Thermal ($\tau_{th}$) and quantum ($\tau_q$) nucleation time of the 
Q*-phase (with B = 85 MeV/fm$^3$) in $\beta$-stable neutrino-free hadronic matter 
as a function of temperature at fixed pressure $P= 57$~MeV/fm$^3$.  
The crossover temperature is  $T_{co}=7.05$~MeV. 
The limiting conversion temperature for the proto-hadronic star is, in this case, $\Theta = 10.3$~MeV, 
obtained from the intersection of the thermal nucleation time curve (continuous line) and the 
dot-dashed line representing  $\log_{10}(\tau/{\rm s}) = 3$.}  
\label{Fig6}
\end{figure}
%______________________________________________________________
%%%%%%%%%%%%%%%%%%%%%%%%%%%%%%%%%%%%%%%%%%%%%%%%%

%%%%%%%%%%   TABLE 4   %%%%%%%%%%%%%%%%%%%%%%%
\begin{table}
\begin{tabular}{ccc}
\hline
$P$   & $T_{co}$ &~~$\log_{10}(\tau/{\rm s}) $  \\
\hline
   53.98~~~~&~~  5.0~~~~~&  233.6  \\
   55.48~~~~&~~  6.0~~~~~&  121.3  \\
   56.94~~~~&~~  7.0~~~~~&    56.6  \\
   58.42~~~~&~~  8.0~~~~~&    16.0  \\ 
   58.85~~~~&~~  8.3~~~~~&      3.0   \\
\hline
\end{tabular}
\caption{Crossover temperature $T_{co}$ (in MeV), for different fixed values of the 
pressure $P$ (in Mev/fm$^3$) of neutrino-free hadronic matter. 
The third column reports the logarithm of the nucleation time (in seconds) calculated 
at the crossover temperature. The value $8.3$ MeV defines the value of the 
limiting conversion temperature $\Theta$ for a star with a 
central pressure $P=58.85$ MeV fm$^{-3}$.}  
\label{table4}
\end{table}
%%%%%%%%%%%%%%%%%%%%%%%%%%%%%%%%%%%%%%%%%%%%%%%%

\section{Evolution of proto-hadronic stars}

Keeping in mind the physical conditions in the interior of a proto-hadronic star 
(Burrow \& Lattimer \cite{BurLat86}; Prakash et al. \cite{prak97}) (see Sect. 1 of the present paper)   
to establish whether this star will survive the early stages of its evolution without ``decaying" to 
a quark star, one has to compare the quark matter nucleation time 
$\tau=\min(\tau_q,\tau_{th})$ with the cooling time  $t_{cool} \sim$~a~few~$10^2$~s.   
If $\tau >> t_{cool}$ then quark matter nucleation is not likely to occur 
in the newly formed star, and this star will evolve into a cold deleptonized configuration.  
We thus introduce the concept of {\it limiting conversion temperature} $\Theta$ for the  
proto-hadronic star and define it as the value of the stellar central temperature $T_c$ for which 
the Q*-matter nucleation time is  equal to $10^3$~s. The limiting conversion temperature 
$\Theta$ will clearly depend on the value of the stellar central pressure (and thus on the value 
of the stellar mass).  

%%%%%%%%%%%%%%%%%%%%  FIG. 7   %%%%%%%%%%%%%%%%%%%%%%%%%%%%%%%%
%                                                One column figure
%----------------------------------------------------------- S_vib
\begin{figure}
\centering
\vspace*{18pt}
\includegraphics[width=8cm] {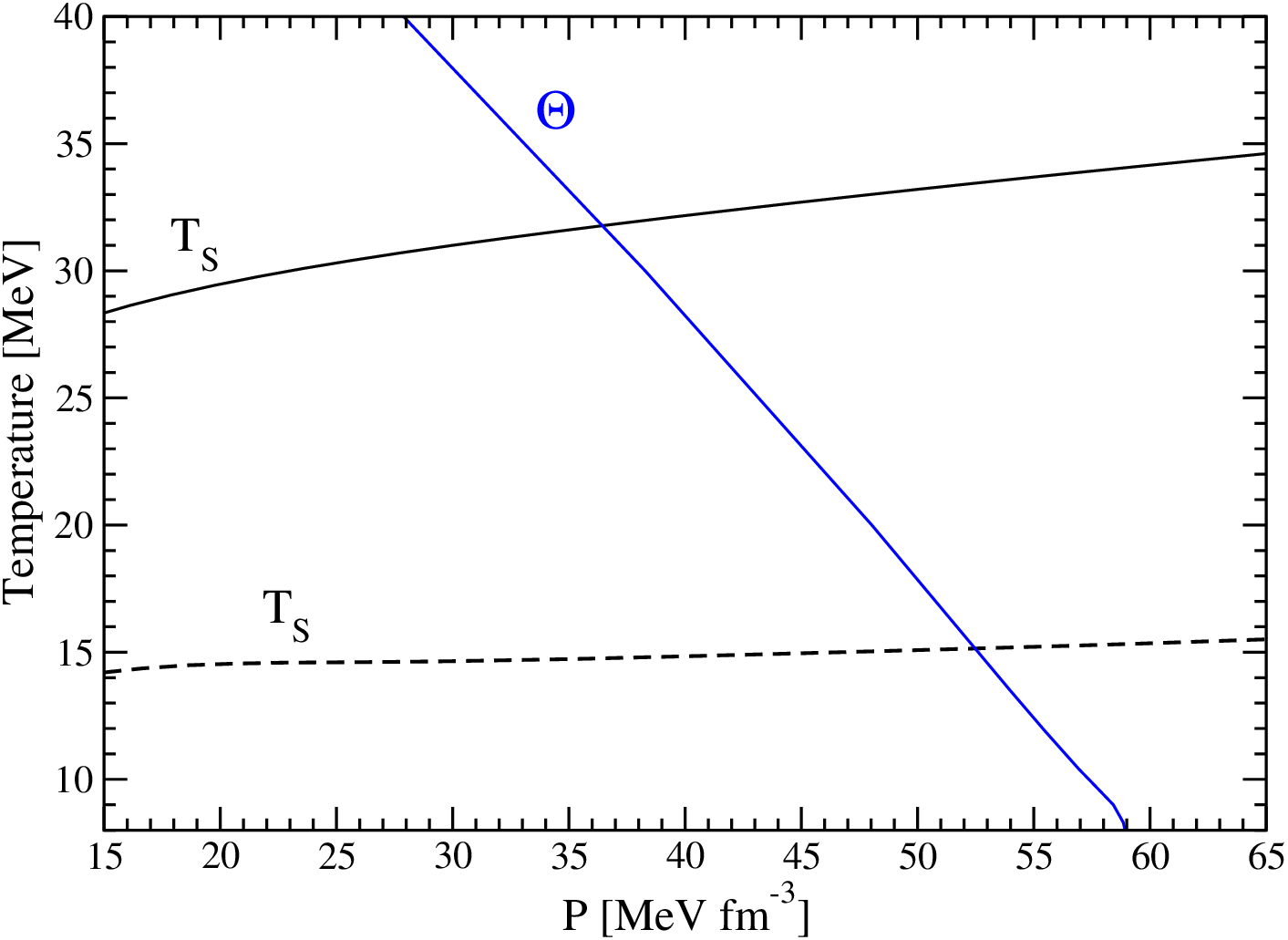}    %{theta_B85.eps}
\caption{(Color online.) The limiting conversion temperature $\Theta$ for a newborn hadronic star 
as a function of the  central stellar pressure. The lines labeled  $T_S$ represent the stellar matter temperature  as a function of pressure at fixed entropies per baryon $\tilde S/k_B = 1$ 
(dashed line) and $2$ (solid line). Results for neutrino-free matter. }   
\label{Fig7}
\end{figure}
%______________________________________________________________

%%%%%%%%%%%%%%%%%%%%  FIG. 8   %%%%%%%%%%%%%%%%%%%%%%%%%%%%%%%%
%                                                One column figure
%----------------------------------------------------------- S_vib
\begin{figure}
\centering
\vspace*{18pt}
\includegraphics[width=8cm] {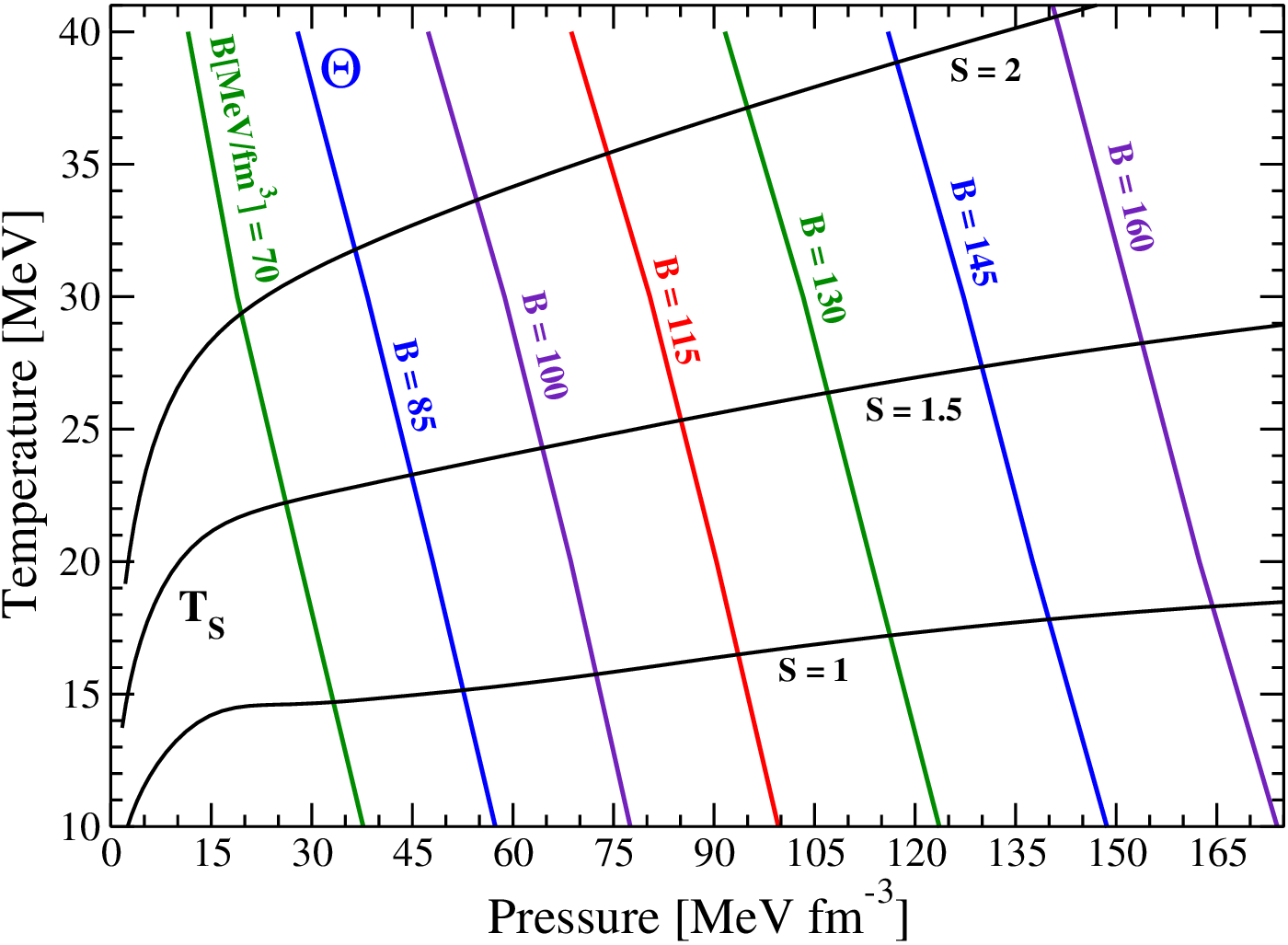}  %{theta_GM1.eps}
\caption{(Color online.) The limiting conversion temperature $\Theta$ for a newborn hadronic star 
as a function of the  central stellar pressure for different values of the bag constant $B$. 
 The lines labeled  $T_S$ represent the stellar matter temperature 
as a function of pressure at fixed entropies per baryon $\tilde S/k_B = 1$, $1.5$, $2$.  
Results for neutrino-free matter.}   
\label{Fig8}
\end{figure}
%______________________________________________________________

%%%%%%%%%%%%%%%%%%%%  FIG. 9   %%%%%%%%%%%%%%%%%%%%%%%%%%%%%%%%
%                                                One column figure
%----------------------------------------------------------- S_vib
\begin{figure}
\centering
\vspace*{18pt}
\includegraphics[width=8cm] {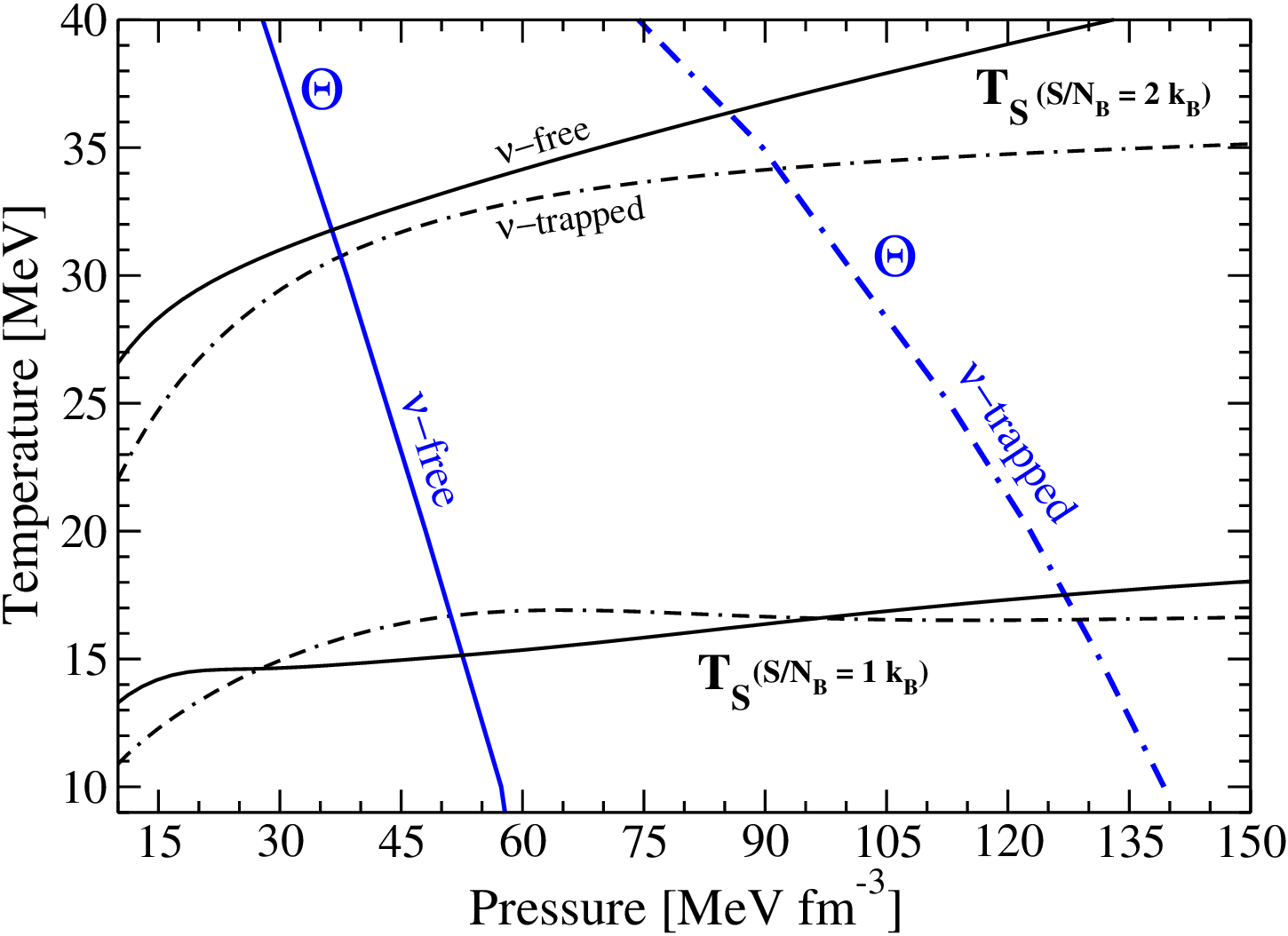}  %{theta_GM1_B85_con_e_senza_neutrini.eps}
\caption{(Color online.) Effects of neutrino trapping on the limiting conversion temperature $\Theta$ 
for a newborn hadronic star as a function of the  central stellar pressure.}
\label{Fig9}
\end{figure}
%______________________________________________________________
%%%%%%%%%%%%%%%%%%%%%%%%%%%%%%%%%%%%%%%%%%%%%%%%%

%%%%%%%%%%%%%%%%%%%%  FIG. 10   %%%%%%%%%%%%%%%%%%%%%%%%%%%%%%%%
%                                                One column figure
%----------------------------------------------------------- S_vib
\begin{figure}
\centering
\vspace*{18pt}
\includegraphics[width=8cm] {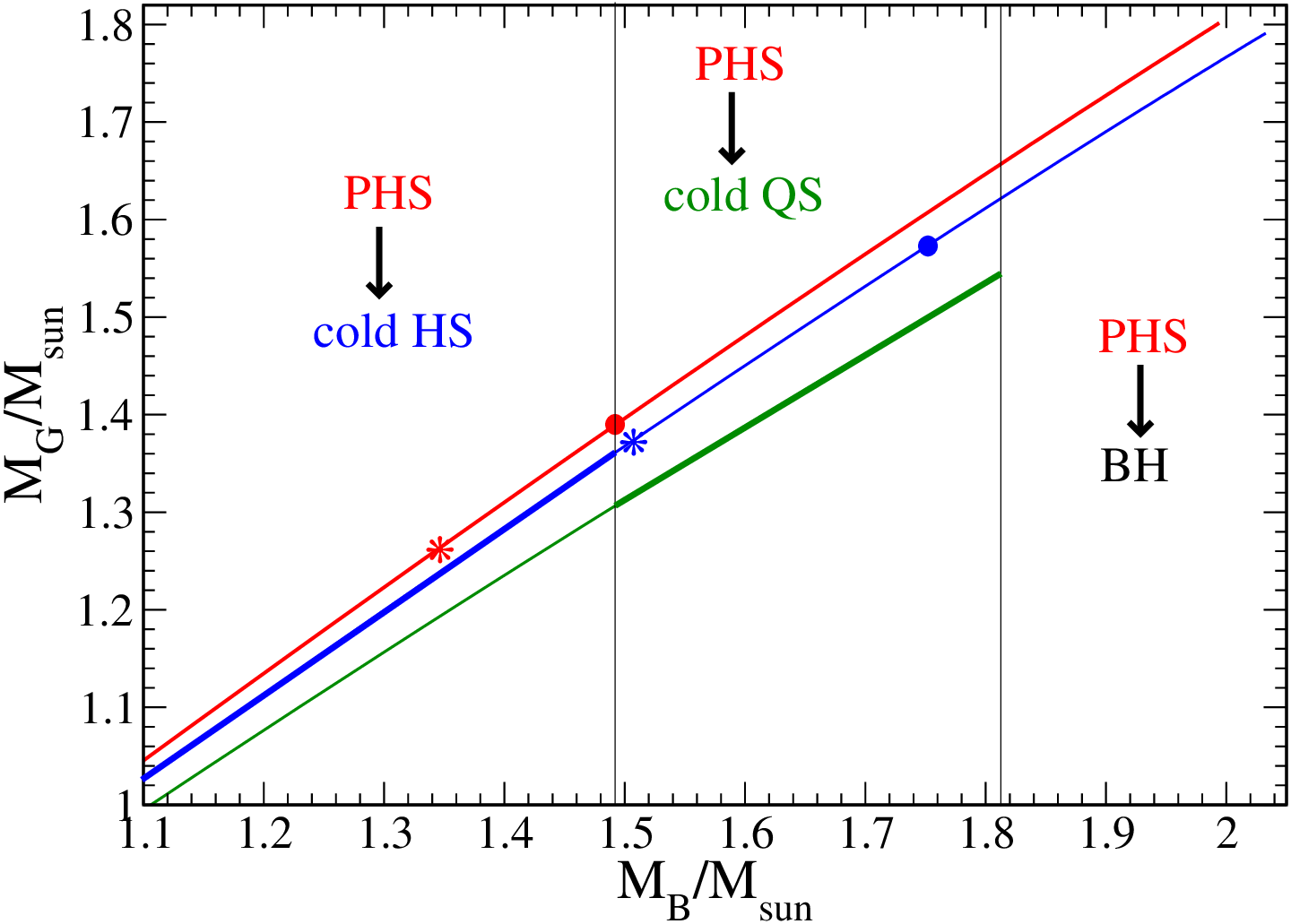}     %{Mg_Mb_GM1_B85_S2.eps}
\caption{(Color online.)  Evolution of a proto-hadronic star in the gravitational--baryonic mass plane for 
B = 85 MeV/fm$^3$ and $\sigma =30$~MeV/fm$^2$. 
The upper (red) line represents the stellar equilibrium sequence for neutrino-free  
proto-hadronic stars (PHS) with $\tilde {S} = 2~k_B$.  
The middle (blue) line represents the cold the HS sequence. 
The asterisk and the full circle on these lines represent the stellar configuration with nucleation 
time $\tau = \infty$  and the critical mass configuration $M_{cr}$, respectively.  
The lower (green) line represent the cold QS sequence.    
Assuming  $M_B =~$const, the evolution of a PHS in this plane occurs along a vertical line.
Stellar masses are in units of the solar mass, $M_{sun} = 1.989 \times 10^{33}$~g.} 
\label{Fig10}
\end{figure}
%______________________________________________________________

%%%%%%%%%%%%%%%%%%%%  FIG. 11   %%%%%%%%%%%%%%%%%%%%%%%%%%%%%%%%
%                                                One column figure
%----------------------------------------------------------- S_vib
\begin{figure}
\centering
\vspace*{18pt}
\includegraphics[width=8cm] {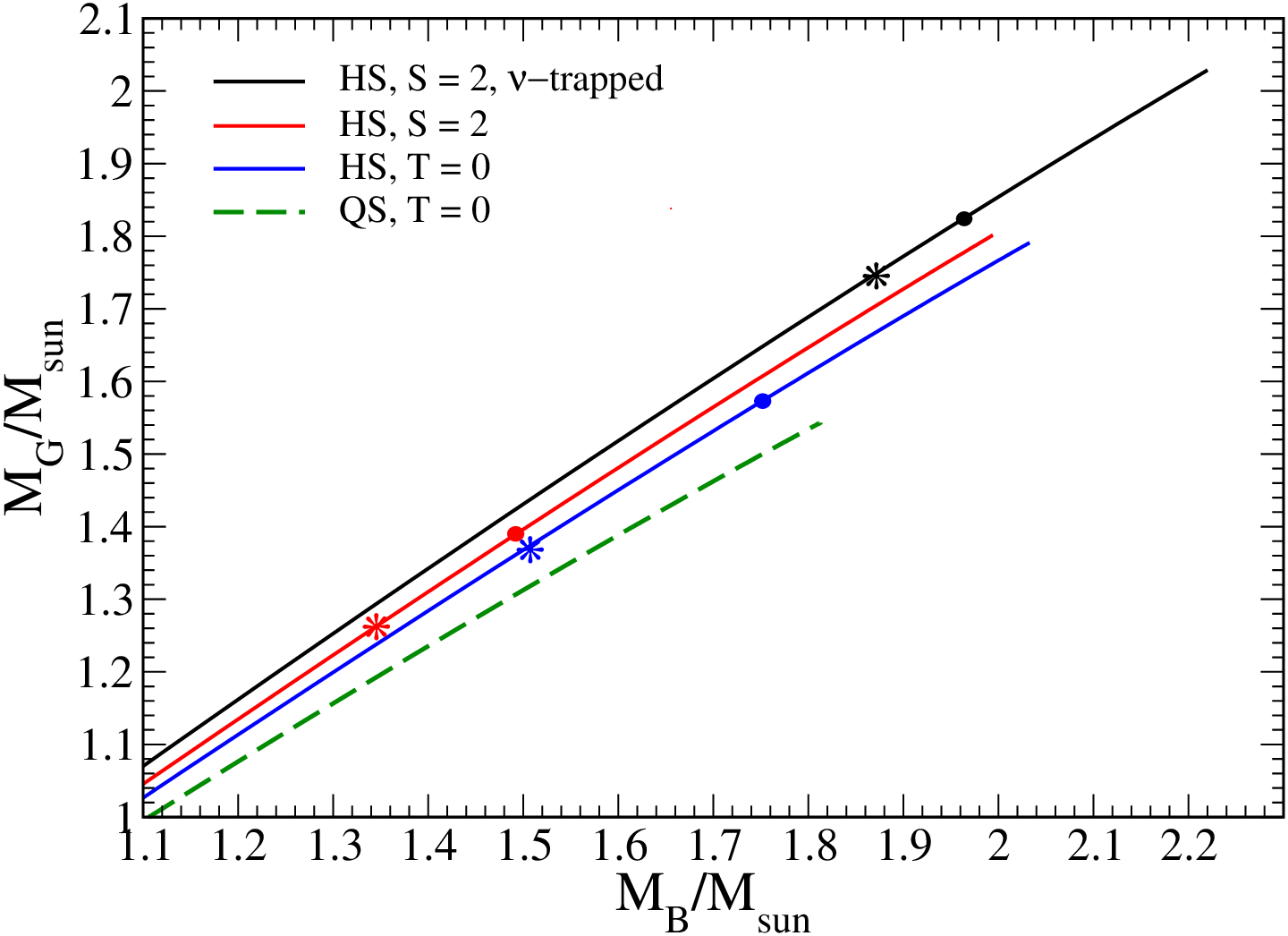}  %{Mg_Mb_85.eps}
\caption{(Color online.) Same as Fig. 10, but with the additional stellar sequence 
(upper (black) line)  for proto-hadronic stars with trapped neutrinos ($\nu$PHSs) with a 
lepton fraction $Y_L = 0.4$ and $\tilde {S} = 2~k_B$. }
\label{Fig11}
\end{figure}
%______________________________________________________________

%%%%%%%%%%%%%%%%%%%%  FIG. 12   %%%%%%%%%%%%%%%%%%%%%%%%%%%%%%%%
%                                                One column figure
%----------------------------------------------------------- S_vib
\begin{figure}
\centering
\vspace*{18pt}
\includegraphics[width=8cm] {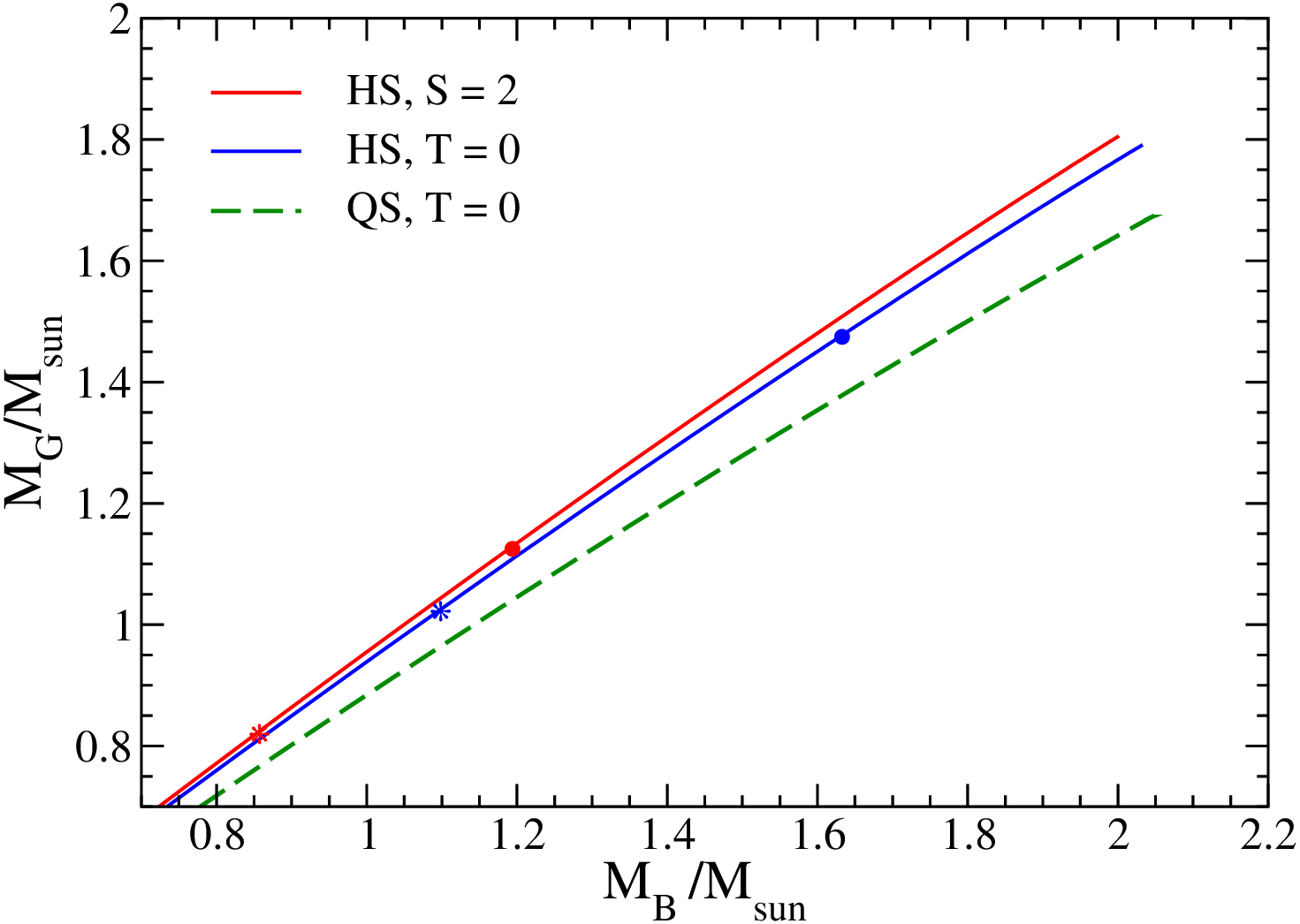}     %{Mg_Mb_70.92.eps}
\caption{(Color online.) Same as Fig. 10, but with a bag constant B = 70.9 MeV/fm$^3$.}
\label{Fig12}
\end{figure}
%______________________________________________________________

The limiting conversion temperature $\Theta$ is plotted in Fig.~7 as a function of the stellar 
central pressure $P$.  A proto-hadronic star with a central temperature  $T_c > \Theta(P)$  
will likely nucleate a Q*-matter drop during the early stages of its evolution,  
will finally evolve to a cold and deleptonized quark star, or will collapse to a black hole 
(depending on the particular model adopted for the matter EOS).  

For an isoentropic stellar core (Burrow \& Lattimer \cite{BurLat86}; Prakash et al. \cite{prak97}),  
the central temperature of the proto-hadronic star is given for the present EOS model  by the lines 
labeled by $T_S$ in Fig.~7,  relative to the case $\tilde {S} = 1~k_B$  and 
$\tilde {S} = 2~k_B$.    
The intersection point ($P_S,\Theta_S$) between the two curves $\Theta(P)$ and $T_S(P)$
thus gives the central pressure and temperature of the configuration that we denote as 
the {\it critical mass} configuration of the proto-hadronic stellar sequence.  
The value of the gravitational critical mass $M_{cr} = M(P_S,\Theta_S)$  and baryonic critical mass 
$M_{B,cr}$ are reported in Table~5  for three different choices of the entropy per baryon,  
$\tilde {S}/k_B = 0$ (corresponding to a cold hadronic star)
%********
{\footnote 
{In Ref. (Berezhiani et al. \cite{be02}; \cite{be03}; Bombaci et al. \cite{bo04}; 
Drago et al. \cite{drago04}; Lugones \& Bombaci \cite{lug05}; Bombaci et al. \cite{blv07}; 
Bombaci et al. \cite{bppv08}),  the critical mass for cold ($T=0$) 
metastable hadronic stars has been defined as the value of the gravitational mass for 
which the quantum nucleation time is equal to one year:  $M_{cr}(T=0) = M(\tau_q=1~{\rm yr})$.   
It is worth recalling that the nucleation time $\tau_q$ is an extremely steep function of the 
hadronic star mass (Berezhiani et al. \cite{be02}; \cite{be03}; Bombaci et al. \cite{bo04}), therefore 
the exact value of $\tau_q$ chosen in the definition of  $M_{cr}(T=0)$ is not crucial (one must take a  ``reasonable low'' value of $\tau_q$, much shorter than the age of young pulsars as the Crab pulsar). 
We have verified that changing  $\tau_q$ from 1~yr to $10^3$~s modifies $M_{cr}(T=0)$ by $\sim 0.02\%$.    
On the other hand, the nucleation time $\tau=\min(\tau_q, \tau_{th})$ entering in the definition 
of the critical mass of proto-hadronic stars $M_{cr}(\tilde{S})$ must be comparable to the  
proto-hadronic star cooling time $t_{cool}$.  
}}  
%*********************
, $1$ and $2$ in the case of a bag constant B = 85 MeV/fm$^3$ and for $\sigma = 30$~MeV/fm$^2$.  
In the same table, we also report the value of the gravitational mass ${\cal M}$ of the 
cold hadronic star with baryonic mass equal to $M_{B,cr}$.     
This configuration is stable ($\tau=\infty$) with respect to Q*-matter nucleation in the case   
$\tilde {S}/k_B = 2$, and it is essentially stable (having a nucleation time enormously longer 
than the age of the universe) in the case $\tilde {S}/k_B = 1$.  

%%%%%%%%%%   TABLE 5   %%%%%%%%%%%%%%%%%%%%%%%
\begin{table}
\begin{tabular}{cccc}
\hline
  $\tilde {S}/k_B$ & $M_{cr}$  & $M_{B,cr}$ & ${\cal M}$ \\
\hline
   0.0~~~~~  & 1.573~~~~~ & 1.752~~~~~ & 1.573 \\   
   1.0~~~~~  & 1.494~~~~~ & 1.643~~~~~ & 1.485 \\   
   2.0~~~~~  & 1.390~~~~~ & 1.492~~~~~ & 1.361 \\
\hline
\end{tabular}
\caption{Gravitational ($M_{cr}$) and baryonic ($M_{B,cr}$) critical mass (see text for more details) 
for proto-hadronic stars at different entropy per baryon  $\tilde {S}/k_B$.  
${\cal M}$ denotes the gravitational mass of  the cold hadronic configuration with the same 
stellar baryonic mass ($M_{B,cr}$).  
Stellar masses are in units of the solar mass, $M_{sun} = 1.989 \times 10^{33}$~g.
Results are relative to a bag constant  B = 85 MeV/fm$^3$}   
\label{table5}
\end{table}
%%%%%%%%%%%%%%%%%%%%%%%%%%%%%%%%%%%%%%%%%%%%%%%%    

The limiting conversion temperature $\Theta$ for a newborn hadronic star is plotted in Fig.~8 
for different values of the bag constant. The increase in B produces a growth of the region of 
the P-T plane  where the proto-hadronic star could survive Q* nucleation and thus evolve to a cold 
hadronic star.  

To explore the role of neutrino trapping on the limiting conversion temperature $\Theta$ and on the 
critical mass of the proto-hadronic star,  we plot in Fig.~9 the results of our calculations for  $\Theta$ 
and for the stellar matter temperature $T_S$ at fixed entropies per baryon  ($\tilde S/k_B = 1$, and  $2$)  
in the case of neutrino-free matter  and neutrino-trapped matter.   
Results in Fig.~9 are relative to the GM1 EOS for the hadronic phase and the bag model EOS with 
B = 85 Mev/fm$^3$ for the quark phase.  As we see, neutrino trapping has a conspicuous effect on  
$\Theta$, and ultimately strongly suppresses the nucleation of Q*-matter in the proto-hadronic star. 
There is also a relatively small effect of neutrino trapping on $T_S$.  
 
The evolution of a proto-hadronic star (PHS) within our scenario is delineated in Fig.~10, where  
we plot the appropriate stellar equilibrium sequences in the gravitational--baryonic mass plane 
obained from the GM1 EOS for the hadronic phase and the bag model EOS with B = 85 MeV/fm$^3$ 
for the quark phase.    
In particular, we plot the PHS sequence, {\it i.e.} isoentropic HSs ($\tilde {S} = 2~k_B$) 
and neutrino-free matter (upper line), and the cold HS sequence (middle line). 
The asterisk and the full circle on these lines identify respectively the stellar configuration with 
$\tau = \infty$ and the critical mass configuration.  
We denote as $M_{B,cr}^{PHS} \equiv M_{B,cr}(\tilde S = 2 k_B)$ the baryonic critical mass for the 
PHS sequence and as $M_{B,cr}^{HS} \equiv M_{B,cr}(\tilde S = 0)$ the baryonic critical mass for the 
cold hadronic star sequence.  
Finally, the lower line represents the cold QS sequence. 
We assume (Bombaci \& Datta \cite {grb})  $M_B =$~const during these stages of the stellar evolution
)
%********
{\footnote 
{Sizeable mass accretion on the proto-neutron star occurs within a time of $\sim 0.5$~s after core bounce.
(Burrow \& Lattimer \cite{BurLat86}; Prakash et al. \cite{prak97}). During the subsequent stages,  
the star thus evolves with $M_B \simeq$~const.  
}}.  
%********************* 
Thus according to the results in Fig.~10,  proto-hadronic stars with a baryonic mass  
$M_B < M_{B,cr}^{PHS}$ ($=1.492~M_{sun}$ within the present EOS parametrzation)  
will survive Q*-matter {\it early nucleation}  ({\it i.e.} nucleation within the cooling time  
$t_{cool} \sim$~a~few~$10^2$~s) and in the end will form stable ($\tau = \infty$) cold hadronic stars.    
Proto-hadronic stars with $M_{B,cr}^{PHS} \le  M_B < M_{B,max}^{QS}$ (the maximum baryonic mass  
of the cold QS sequence, $1.813~M_{sun}$ for the present EOS) will experience early nucleation 
of a Q*-matter drop and will ultimately form a cold deleptonized quark star.  
The last possibility is for PHSs having  $M_B >  M_{B,max}^{QS}$. In this case the early nucleation 
of a Q*-matter drop will trigger a stellar conversion process  to a cold QS configuation with 
$M_B > M_{B,max}^{QS}$, thus these PHSs will finally form  black holes. 

The outcomes of this scenario are not altered by neutrino trapping effects in hot $\beta$-stable 
hadronic matter as illustrated in Fig.~11 where (in addition to the stellar sequence curves of Fig.~10)  
we plot the stellar sequence for proto-hadronic stars with trapped neutrinos ($\nu$PHSs) with a 
lepton fraction $Y_L = 0.4$ and $\tilde {S} = 2~k_B$.  
For the $\nu$PHS stellar sequence in Fig.~11 we find a critical baryonic mass  
$M_{B,cr}^{\nu PHS}  = 1.96~M_{sun}$. 
Thus $\nu$PHSs with $M_B \geq M_{B,cr}^{\nu PHS} $ after neutrino escape and cooling will finally 
evolve to black holes.  
The fate of a $\nu$PHS with  $M_B < M_{B,cr}^{\nu PHS}$ is the same as the corresponding 
neutrino-free PHS with equal baryonic mass.  

Finally, in Fig.~12 we plot the PHS, cold HS, and cold QS sequences in the gravitational--baryonic mass 
plane for the case of a bag constant B = 70.9 MeV/fm$^3$. (All the other EOS parameters are the same 
as for the results in Fig. 10.)  Apart from changes in the numerical values of the maximum masses of the various stellar sequences and of the values of the stellar critical masses, the evolutionary scenario of a 
proto-hadronic star is qualitatively similar to the one previously discussed for the results in Fig. 10.   

An interesting situation comes up when a PHS with $M_B < M_{B,cr}^{PHS}$ is formed in a 
binary stellar system. In this case, as we have just seen, a stable (case of Fig. 10) or a 
metastable (case of Fig. 12) cold HS is formed.  Long-term accretion from the companion star will start 
to populate the portion of the cold HS sequence with $M_{B,cr}^{PHS} < M_B < M_{B,cr}^{HS}$. 
If the star accretes enough matter ($\sim$ 0.1 -- 0.3 $M_{sun}$, see Figs. 10 and 12), it could possibly 
reach the critical mass $M_{B,cr}^{HS}$.  At this point a quantum nucleation process will trigger 
the conversion of this critical mass cold HS to a QS (Berezhiani et al. \cite{be02}; \cite{be03}; 
Bombaci et al. \cite{bo04}). 
The time delay between the supernova explosion forming the PHS, and the second ``explosion"  
(neutrino burst or/and GRB) forming the QS  will actually depend on the mass accretion rate, which 
modifies the quark matter nucleation time via an explicit time dependence of the stellar central 
pressure.  

Within the present scenario, there is thus a stellar mass range  $M_{B,cr}^{PHS} < M_B < M_{B,cr}^{HS}$ 
where it is possible to find two different types of ``neutron stars'' in the Universe:  
pure hadronic compact stars with large radii in the range 12 -- 20 km, and quark stars with small 
radii in the range of 6 -- 9 km (Bombaci et al. \cite{bo04}; Drago \& Lavagno \cite{DL10}).  
Accurate measurements of both the mass and radius of a few individual ``neutron stars'' 
(Bhattacharyya 2010; Steiner et al. 2010) could shed light on the validity of this presumptive scenario.  
 
\section{Limiting mass of compact stars}

The possibility of having metastable hadronic stars, together with the feasible existence 
of two distinct families of compact stars (pure hadronic stars and quark stars), 
demands an extension of the concept of maximum mass of a ``neutron star'' with respect 
to the {\it classical} one introduced by Oppenheimer \& Volkoff (1939).     
Since metastable HS with a ``short'' {\it mean-life time} are very unlikely to be observed, 
an extended concept of maximum mass has to be introduced in view of the comparison with the  
values of the mass of compact stars deduced from direct astrophysical observation.   
With this operational definition in mind, Bombaci et al. (\cite{bo04}) define the 
{\it limiting mass} ($M_{lim}$) of a compact star in the case of cold stellar configurations. 
This concept of limiting mass can be straightforwardly extended to the present case, 
{\it i.e.} taking the effects of  proto-neutron stars evolution into account.   
For the two cases reported in Figs. 10 and 12, the limiting mass is the highest between  
the gravitational critical mass  $M_{cr}^{HS}$ of the cold HS sequence and  
the gravitational maximum  mass  $M_{max}^{QS}$ of the cold QS sequence.  
Thus $M_{lim} = M_{cr}^{HS} $ (case in Fig. 10) and  $M_{lim} = M_{max}^{QS} $ (case in Fig. 12). 

  The very recent measurement (Demorest et al. \cite{demo10}) of the Shapiro delay in 
the binary millisecond pulsar J1614-2230 has helped to obtain the mass of the associated 
neutron star. The calculated mass is  $M = (1.97 \pm 0.04) M_\odot$ (Demorest et al. 
\cite{demo10}), making  PSR~J1614-2230 the most massive neutron star known to date.  

As is well known (see {\it e.g.} Lattimer \& Prakash \cite{LP01}; Bombaci \cite{bomb07}),  
neutron star mass measurements give one of the most stringent tests of the composition 
and EOS of strong interacting matter at  very high densities. 
Here, we very briefly address some possible implications of the high mass of PSR~J1614-2230  
in connection with the feasibility of the scenario discussed in present work.  
To this aim we consider two almost diametrical situations for the stiffness of the EOS.   

The first possibility is for the case of a stiff hadronic EOS  and a soft quark matter EOS. 
This is the case, for example, for the GM1 model with $x_\sigma = x_\rho  = 0.8$, 
$x_\omega = 0.913$ for hyperonic matter, and the MIT bag model EOS 
(Farhi \& Jaffe \cite{fj84})) with $B = 100$--150 MeV/fm$^3$ for quark matter.  
For these EOS models, the limiting mass is given by the gravitational critical mass of the HS 
sequence, and it is in the range $M_{lim} = 1.9$--$2.1 M_\odot$ (Bombaci et al. \cite{bppv08}).        
Thus, in this case, PSR~J1614-2230 can be interpreted as a pure hadronic star (hyperon star),  
and quark matter nucleation, in HS with $M > M_{cr}^{HS}$, will produce stellar configurations  
which will form black holes.   
 
The second possibility is for a soft hadronic EOS and a stiff quark matter EOS.
This is the case, for example, of the microscopic Brueckner-Hartree-Fock EOS for hyperonic 
matter (Baldo et al. \cite{bal00}; Vida\~na et al. \cite{vid00}; Schulze et al. \cite{sch06}) 
and the recent perturbative calculations of the quark matter EOS by (Kurkela et al. \cite{kurk10}) which include quark interaction effects up to the second order in the QCD coupling $\alpha_s$ 
(see also Fraga et al. \cite{frag01}; Alford et al. \cite{alf05}).    
For these EOS models, the limiting mass is likely given  by the gravitational maximum mass 
of the cold QS sequence, which according to the results of (Kurkela et al. \cite{kurk10})  
is $M_{max}^{QS} \sim 2 M_\odot$ (hybrid stars), $M_{max}^{QS} = 2.0$--$2.7 M_\odot$ 
(strange stars). Thus, for this second case, PSR~J1614-2230 could  be interpreted as 
a quark star (\"Ozel et al. \cite{ozel10}).

\section{Conclusions}

In summary, in this work we have studied the quark deconfinement phase transition in hot 
$\beta$-stable hadronic matter and explored some of its consequences for the physics 
of neutron stars at birth.  
We calculated and compared the nucleation time due to thermal and quantum nucleation mechanisms,  
and computed the crossover temperature above which thermal nucleation dominates the finite  
temperature quantum nucleation mechanism. 
In addition, we introduced the new concept of {\it limiting conversion temperature} 
$\Theta$ for proto-hadronic stars and  extended the concept of critical mass 
(Berezhiani et al. \cite{be02}; \cite{be03}; Bombaci et al. \cite{bo04}) to the case of finite temperature hadronic stars.    

Our main finding is that proto-hadronic stars with a gravitational mass lower than the critical mass 
$M_{cr}$ could survive the early stages of their evolution without decaying to a quark star.  
This outcome contrasts with the predictions of the earlier studies (Horvath et al. \cite{ho92}; 
Horvath \cite{ho94}; Olesen \& Madsen  \cite{ol94}; Heiselberg \cite{hei95}; Harko et al.  \cite{harko04})    
where it was inferred that all the pure hadronic compact stars, with a central temperature above 
2 -- 3 MeV,  are converted to quark stars within the first seconds after their birth. 

However, the prompt formation of a critical size drop of quark matter could take place when 
$M > M_{cr}$.  These proto-hadronic stars evolve to cold and deleptonized quark stars  
or collapse to a black holes.  

Finally, if quark matter nucleation occurs during the post-bounce stage of core-collapse supernova, 
then the quark deconfinement phase transition could trigger a delayed supernova explosion 
characterized by a peculiar neutrino signal (Sagert et al. \cite{sage+09}; Mintz et al. \cite{mintz10}; 
Nakazato et al. \cite{Naka08}; Dasgupta et al. \cite{dasg10}).

%%%%%%%%%%%%%%%%%%%%%%%%%%%%%%%%%%%%%%%%%%%%%%%%%%%%%%
\section*{Acknowledgement}  
This work has been partially supported by FCT (Portugal) under grants SFRH/BD/62353/2009,  
PTDC/FIS/64707/2006  
and FCOMP-01-0124-FEDER-008393 with FCT reference CERN/FP/109316/2009,  
and by COMPSTAR, an ESF Research Networking Programme.  
%%%%%%%%%%%%%%%%%%%%%%%%%%%%%%%%%%%%%%%%%%%%%%%%%%%%%%

%%%%%%%%%%%%%%%%%%%%%%%%%%%%%%%%%%%%%%%%%%%

%%%%%%%%%%%%%%%%%%%%%%%%%%%%%%%%%%%%%%%%%%%%%%%%%%%%%%%%

\begin{thebibliography}{}
%%%%%%%%%%%%%%%%%%%%%%%%%%%%%%%%%%%%%%%%%%%%

\bibitem[2004]{ABG04}   Aguilera, D. N., Blaschke, D., \& Grigorian, H.  2004, A\&A, 416, 991.   

\bibitem[2005]{alf05}  Alford, M. G., Braby, M., Paris, M., \& Reddy, S. 2005, ApJ, 629, 969.

\bibitem[2008]{alf+08}  Alford, M. G.,  Schmitt, A., Rajagopal, K., \& Schafer, T.   
                                   2008, Rev. Mod. Phys.  80, 455.

\bibitem[2000]{bal00}   Baldo, M.,  Burgio, G. F., \&  Schulze, H.-J. 
                                         2000, Phys. Rev. C, 61, 055801. 

\bibitem[2008]{bambi08} Bambi, C., \& Drago, A. 2008, Astropart. Phys. 29, 223. 
\bibitem[2007]{Benh07}  Benhar,  O., \& Valli, M. 2007, Phys. Rev. Lett., 99,  232501.
\bibitem[2010]{Benh10}   Benhar, O., Polls, A., Valli, M. \&  Vida\~na, I.  2010,  Phys. Rev. C, 81, 024305. 

\bibitem[2002]{be02} Berezhiani, Z., Bombaci, I.,  Drago, A.,  Frontera, F., \& Lavagno, A.  
                                 2002, Nuclear Physics B - Proceedings Supplements, 113, 268  
\bibitem[2003]{be03} Berezhiani, Z., Bombaci, I.,  Drago, A.,  Frontera, F., \& Lavagno, A. 
                                2003, ApJ, 586, 1250.
\bibitem[2010]{Bhat2010}  Bhattacharyya, S.  2010, Adv. Space Res., 45, 949.  
\bibitem[1971]{bod71}     Bodmer, A. R. 1971, Phys. Rev. D 4,1601. 
\bibitem[1996]{bomb96} Bombaci, I. 1996, A\&A  305,  871. 
\bibitem[2007]{bomb07} Bombaci, I. 2007, Eur. Phys. J.  A 31, 810. 
\bibitem[2000]{grb}       Bombaci, I. \& Datta, B. 2000, ApJ 530,  L69.
\bibitem[2004]{bo04}     Bombaci, I.,  Parenti, I., \&  Vida\~na, I. 2004, ApJ 614, 314.
\bibitem[2007]{blv07}    Bombaci, I., Lugones, G., \&  Vida\~na, I. 2007,  A\&A  462, 1017.
\bibitem[2008]{bppv08}  Bombaci, I., Panda, P.K., Provid\^encia, C., \& Vida\~na, I. 
                                     2008, Phys. Rev. D 77, 083002.
\bibitem[2009]{blppv09}  Bombaci, I., Logoteta, D., Panda, P.K., Provid\^encia, C., \& Vida\~na, I. 
                                     2009, Phys. Lett. B 680, 448.

\bibitem[1986]{BurLat86}  Burrows, A., \& Lattimer, J. M. 1986, ApJ, 307,  178.

\bibitem[2004]{CN04}   Casalbuoni, R., \& Nardulli, G. 2004, Rev. Mod. Phys. 76,  263.
\bibitem[2006]{CB06} Chattaerjee, D. \& Bandyopadhyay, D. 2006, Phys. Rev. D, 74, 023003. 
\bibitem[1992]{CseKap92}  Csernai, L., \&  Kapusta, J. I.  1992, Phys. Rev. D, 46, 1379.

\bibitem[1984]{Dan84}  Danielewicz, P. 1984, Phys. Lett. B, 146, 168. 
\bibitem[2010]{dasg10}  Dasgupta, B., Fischer, T., Horiuchi, S., Liebend\"orfer, M., \& 
                                    Mirizzi, A.  2010, Phys. Rev. D, 81, 103005.

\bibitem[2010]{demo10} Demorest, P. B., Pennucci, T., Ransom, S. M., Roberts, M. S. E.   \& 
                                    Hessel, J. W. T.  2010, Nature, 467, 1081.

 
\bibitem[2004]{drago04} Drago, A., Lavagno, A., \& Pagliara, G. 2004, Phys. Rev. D 69, 057505.
\bibitem[2010]{DLP07}   Drago, A., Lavagno, A., \& Parenti, I.  2007, ApJ, 659, 1519.  
\bibitem[2010]{DL10}     Drago, A., \& Lavagno, A.  2010,  arXiv:1004.0325. 

\bibitem[2008]{dps-b08} Drago, A., Pagliara, G., \& Schaffner-Bielich, J. 2008, J. Phys. G 35, 014052. 

\bibitem[1984]{fj84}   Farhi, E., \& Jaffe, R. L. 1984,  Phys. Rev. D 30,  2379.  
\bibitem[2004]{fk04}  Fodor, Z.,  \&   Katz, S. D. 2004, Prog. Theor.  Suppl., 153, 86. 
\bibitem[1979]{FI79}   Flowers, E. \& Itho, N. 1979,  ApJ, 230, 847.

\bibitem[2001]{frag01}  Fraga, E. S., Pisarski, R. D. \&  Schaffner-Bielich, J. 2001, 
                                          Phys. Rev. D, 63, 121702(R). 

\bibitem[2000]{glen00} Glendenning, N. K.  2000, Compact Stars: Nuclear Physics, 
                               Particle Physics, and General Relativity, Springer, New York. 
\bibitem[2001]{gm91} Glendenning, N. K., \&  Moszkowski, S. 1991,  Phys. Rev. Lett., 67, 2414 
\bibitem[2008]{Gu08}   Gu, J.-F., Guo, H., Lee, X.-G,  Liu Y.-X., \& Xu, F.-R.  
                                  2008, Commun. Theor. Phys. (Beijing, China), 49, 461.

\bibitem[2004]{harko04}  Harko, T.,  Cheng, K. S., \& Tang, P. S. 2004, ApJ 608,  945.
\bibitem[1991]{hbp91} Heiselberg, H., Baym, G., \&  Pethick, C. J.  1991, 
                                  Nucl. Phys. B (Proc. Suppl.)  24, 144.
\bibitem[1993]{hei93}  Heiselberg, H., Pethick,  C. J., \& Staubo, E. F. 1993, Phys. Rev. Lett. 70, 1355.
\bibitem[1995]{hei95}  Heiselberg, H. 1995, in Strangeness and Quark Matter, Ed. G. Vassiliadis, 
                           World Scientific, p.  338; arXiv:hep-ph/9501374. 
\bibitem[2008]{HMcP08} Hidaka, Y., McLerran, L., \&  Pisarski, R. D. 2008,  Nucl. Phys.  A 808, 117.  
\bibitem[1992]{ho92}   Horvath, J. E., Benvenuto, O. G., \&Vucetich, H.  1992, Phys. Rev. D 45, 3865.
\bibitem[1994]{ho94}   Horvath, J. E., 1994,  Phys. Rev. D 49,  5590.
\bibitem[1998]{hs98} Hsu, S. D. H.,  \& Schwetz, M. 1998, Phys. Lett. B, 432, 2003 

\bibitem[1997]{iida97}  Iida, K., \& Sato, K.  1997, Prog. Theor. Phys., 98, 277  
\bibitem[1998]{iida98}  Iida, K., \& Sato, K. 1998, Phys. Rev. C  58, 2538.

\bibitem[2005]{karsch05} Karsch, F.  2005,  J. Phys. G31, S633.

\bibitem[2010]{kurk10} Kurkela, A., Romatschke, P., \&  Vuorinen, A. 2010, 
                                         Phys. Rev. D, 81, 105021. 

\bibitem[1968]{lang68}  Langer, J. S. 1968, Phys. Rev. Lett., 21, 973.
\bibitem[1969]{lang69}   Langer, J. S.  1969, Ann. Phys. (N.Y.),  54,  258. 
\bibitem[1973]{LanTur73} Langer, J. S., \& Turski, L. A. 1973,  Phys. Rev. A, 8, 3230.

\bibitem[2001]{LP01}  Lattimer, J. M.  \&  Prakash, M. 2001,  ApJ, 550, 426. 

\bibitem[1972]{lk72}   Lifshitz, I. M., \&  Kagan, Y.  1972, Sov. Phys. JETP,  35, 206.   
\bibitem[2007]{lomb07}   Lombardo, M. P.  2007, Mod. Phys. Lett. A, 22, 457.    
\bibitem[1998]{lb98} Lugones, G., \& Benvenuto, O. G. 1998, Phys. Rev. D 58, 083001.
\bibitem[2005]{lug05} Lugones, G., \& Bombaci, I. 2005, Phys. Rev. D 72, 065021.
\bibitem[2009]{lugones09} Lugones, G., Grunferld, A. G., Scoccola N. N. \& Villavicencio, C. 
                                       2009, Phys. Rev. D 80, 045017.
\bibitem[2009]{lugones10} Lugones, G.,  do Carmo, T. A. S.,  Grunferld, A. G., \& 
                                       Scoccola N. N. 2010, Phys. Rev. D 81, 085012.

\bibitem[2007]{McP07}  McLerran, L., \&  Pisarski, R. D. 2007,  Nucl. Phys.  A 796, 83.
\bibitem[2003]{mp03} Menezes, D. P., \& Provid\^encia, C.  2003, Phys. Rev. C, 68, 035804

\bibitem[2010]{mintz10}  Mintz, B. W., Fraga, E. S. Pagliara, G, \& Schaffner-Bielich, J. 
                                     2010, Phys. Rev. D  81, 123012.

\bibitem[2008]{Naka08}  Nakazato, K., Sumiyoshi, K., \& Yamada, S. 2008, Phys. Rev. D  77, 103006.

\bibitem[1994]{ol94}    Olesen,  M. L.,  \& Madsen, J. 1994, Phys. Rev. D  49,  2698.
\bibitem[1987]{oli87}   Olinto,  A. V. 1987, Phys. Lett. B 192, 71.
\bibitem[1939]{ov39}    Oppenheimer,  J.R., \& Volkoff, G.M. 1939, Phys. Rev., 55, 374  

\bibitem[2010]{ozel10}  \"Ozel, F., Psaltis, D., Ransom, S., Demorest, P.,  \& 
                                          Alford, M. 2010, ApJ, 724, L199. 

\bibitem[1997]{prak97} Prakash, M.,  Bombaci, I., Prakash, M.,  Ellis, P. J.,   
            Lattimer, J. M.,  \& Knorren, R. 1997,  Phys. Rep.,  280, 1 
\bibitem[1999]{pons99}   Pons, J. A., Reddy, S., Prakash, M.,  Lattimer, J. M.,  
             \& Miralles, J. A. 1999, ApJ, 513, 780   

\bibitem[2009]{sage+09}  Sagert, I., Fischer, T., Hempel, M., Pagliara, G., Schaffner-Bielich, J.,  
              Mezzacappa, A., Thieleman, F.-K., \&  Liebend\"orfer, M. 2009, Phys. Rev. Lett. 102, 081101.

\bibitem[2007]{SB07}   Sandin, F.,  \& Blaschke, D.  2007, Phys. Rev. D, 75, 125013.   

\bibitem[2006]{sch06} Schulze, H.-J., Polls, A., Ramos, A.,  \& Vida\~na, I.,
                                        2006, Phys. Rev. C, 73, 058801. 

\bibitem[1994]{Sedr94}  Sedrakian, A. D., Blaschke, D., R\"opke, G. \& Schultz, H. 1994, 
                                    Phys. Lett. B,  338. 111. 
\bibitem[2010]{SLB10}  Steiner,  A. W., Lattimer, J. M., \& Brown E. F., 2010, arXiv:1005.0811. 

\bibitem[1986]{sw86}  Serot, B. D., \& Walecka, J. D.  1986,  Adv. Nucl. Phys.,  16, 1.  

\bibitem[1980]{TurLan80} Turski,  L. A., \& Langer, J. S. 1980, Phys. Rev. A,  22, 2189.

\bibitem[2004]{vDD04}  van Dalen,  E. N. E., \&  Dieperink, A. E. L. 2004, Phys. Rev. C, 69, 025802.
\bibitem[1994]{VenVis94}  Venugopalan, R., \& Vischer, A. P. 1994, Phys. Rev. E, 49,  5849.
\bibitem[2005]{vbp05} Vida\~na, I.,  Bombaci, I., \& Parenti, I. 2005,  Jour. Phys. G, 31, S1165.  

\bibitem[2000]{vid00} Vida\~na, I., Polls, A., Ramos, A., Engvik, L., \& 
                                        Hjorth-Jensen, M.  2000, Phys. Rev. C, 62, 035801. 

\bibitem[1974]{wal74}  Walecka, J. D.  1974, Ann. Phys. (N.Y.),  83, 491  
\bibitem[1984]{witt84}  Witten, E. 1984, Phys. Rev. D, 30, 272.  

\bibitem[2008]{Yakov08}    Shternin, P. S., \& Yakovlev, D. G. 2008, Phys. Rev. D,  78, 063006. 

\bibitem[2010]{ZLZ10}   Zhang, H. F., Lombardo, U. \& Zuo, W. 2010, Phys. Rev. C,  82, 015805. 
%%%%%%%%%%%%%%%%%%%%%%%%%%%%%%%%%%%%%%%%%%%%
\end{thebibliography}
\end{document}